%% file: mg_computer_vision.tex
\documentclass[usenatbib]{mnras}
\usepackage[utf8]{inputenc}
\usepackage{amsmath}
\usepackage{amssymb}
\usepackage{graphicx}
\usepackage{blkarray}
\input{journal_def.tex}

\providecommand{\adsurl}[1]{\href{#1}{ADS}}

\title[Halo Computer Vision]
  {Characterising dark matter haloes with computer vision}
    \author[J.~Merten et al.]{
  Julian Merten $^{1}$
\thanks{\href{julian.merten@physics.ox.ac.uk}{julian.merten@physics.ox.ac.uk}},
   Quim Llorens $^{2,1}$
   and Hans Winther $^{3}$
  \newauthor \\
  $^1$Department of Physics, University of Oxford, Keble Road, Oxford OX1 3RH,
U.K.\\
  $^2$Facultat de F\'{\i}sica, Universitat de Barcelona, Mart\'{\i} i 
Franqu\`{e}s, 1. E-08028 Barcelona, Catalonia, Spain \\
  $^3$Institute of Cosmology \& Gravitation, University of 
Portsmouth, Portsmouth, Hampshire, PO1 3FX, U.K.\\
  }

  \date{\textcopyright~2017 University of Oxford. All rights reserved.\\
  Submitted to the Monthly Notices of the Astronomical Society.}

\begin{document}
\label{firstpage}
\maketitle
\begin{abstract}
This work explores the ability of computer vision algorithms to characterise
dark matter haloes formed in different models of structure formation. We 
produce surface mass density maps of the most massive
haloes in a suite of eight numerical simulations, all based on the same initial
conditions, but implementing different models of gravity. This suite includes a
standard $\Lambda$CDM model, two variations of $f(R)$-gravity, two variations
of Symmetron gravity and three Dvali, Gabadadze and Porrati (DGP) models. We
use the publicly available \texttt{WND-CHARM} algorithm to extract 2919 image
features from either the raw pixel intensities of the maps, or from a variety
of image transformations including Fourier, Wavelet, Chebyshev and Edge
transformations. After discarding the most degenerate models, we achieve more 
than 60\% single-image
classification success rate in distinguishing the four different models of
gravity while using a simple weighted neighbour distance (WND) to define our
classification metric. This number can be increased to more than 70\% if
additional information, such as a rough estimate of the halo mass, is included. 
We find that the classification success steeply
declines when the noise level in the images is increased, but that this trend
can be largely reduced by smoothing the noisy data. 
We find Zernike moments of the
Fourier transformation of either the raw image or its Wavelet transformation to
be the most descriptive feature, followed by the Gini coefficient of several
transformations and the Haralick and Tamura textures of the raw pixel data 
eventually pre-processed by an Edge transformation. The proposed methodology is 
general and does not only
apply to the characterisation of modified gravity models, 
but can be used to
classify any set of models which show variations in the 2D morphology of their
respective structure.
\end{abstract}

\begin{keywords}
 dark matter -- large-scale structure of Universe -- gravitation -- methods: 
numerical.
\end{keywords}


\section{Introduction}
The widely accepted standard model of cosmology is based on general relativity,
a cold dark matter component and a cosmological constant. While this 
$\Lambda$CDM
model is very successful in describing observations of different kinds on
very large scales 
\citep{Betoule2014,Anderson2014,PlanckCollaboration2016,Hildebrandt2017}, 
tensions  between theoretical predictions and observations
arise when entering the realm of non-linear structure formation, at the nexus of
cosmology and astrophysics. These problems include the counted number of galaxy
clusters as a function of mass and redshift \citep{PlanckCollaboration2016a},
the exact shape of the dark matter density profile on scales ranging from
galaxy clusters to dwarf galaxies \citep{Newman2013,Read2016,Walker2011}, the
scale and distribution of halo substructure
\citep{Kauffmann1993,Boylan-Kolchin2012,Schwinn2017}, the diversity in
simulated dark matter profiles \citep{Oman2015} and the unexpected correlation
between baryonic and dark matter components of galaxies
\citep{McGaugh2016}. A thorough understanding of these tensions is difficult
due to the non-linear nature of structure formation. In this regime,
theoretical predictions generally rely on the results of cosmological N-body
and hydrodynamical simulations.

Several models of structure formation have been proposed to remedy some or all
of the aforementioned tensions between observations and the simplest
$\Lambda$CDM simulations. Such models include, but are not limited to, a more
detailed modelling of baryonic effects including feedback processes on
cosmological scales via sub-grid physics \citep[among
others,][]{Vogelsberger2014, Dubois2014, Schaye2015,McCarthy2017} and in
greater detail with smaller boxes using  pristine resolution
\citep{Hopkins2014,Kimm2015,Nelson2016}; alternative models of gravity
\citep[e.g.][and references therein]{Clifton2012}; and more general models of 
dark matter, e.g.~warm dark
matter \citep[e.g.][and references therein]{Lovell2014,Bozek2016} or
self-interacting dark matter \citep{Rocha2013,Peter2013,Vogelsberger2014}.

This recent emergence of a larger variety of structure formation models, poses 
the immediate challenge of finding
their signatures in astrophysical data. In this
work, we aim for characterising the distribution of matter in 
a very agnostic fashion. While there are many
possible descriptors to characterise structure in simulations or real data, our
approach is driven by the fact that gravitational lensing \citep[see][for a
review]{Bartelmann2010a} has recently made great progress in delivering a
detailed and robust 2D picture of the matter distribution in observed massive 
haloes
\citep[e.g.][for recent 
analyses]{Johnson2014a,Merten2015,Jauzac2015,Massey2015,Treu2016,Caminha2016,
Diego2016}. In general, the
characterisation of an observed halo is reduced to a single number such as its
mass, or to a two or three parameter density profile under the assumption of
spherical symmetry. But the increased number of background galaxies for weak
lensing studies 
\citep[e.g.][]{Clowe2006,Massey2007,Merten2011,Dawson2012,Jee2014a,Umetsu2014, 
Melchior2015,Harvey2015,Medezinski2016} combined with more sophisticated 
modelling
techniques, which combine multi-scale tracers such as weak -and strong lensing 
\citep[e.g.][]{Bradav2009,Jullo2009,Merten2009,Diego2015,Merten2016},
allows for a detailed reconstruction of the total matter distribution on
a range of scales. The full morphological
information in such 2D maps shall be used to draw conclusion on the mechanisms 
of structure
formation. This approach avoids the usual compression of valuable information
into very  few numbers and hence stands a better chance of being able to
distinguish the potentially subtle differences between
structure formation models.

The scope of this work is a pilot study of the mentioned approach in the
context of modified gravity models, but the framework is general and can be
used to characterise haloes in any model of structure formation. Our approach
uses computer vision techniques to extract the morphological information in
surface mass density maps of dark matter haloes. This approach developed from
digital image processing \citep{Gonzalez2007} and is now routinely used in
scientific applications such as medical imaging, microscopy or material
sciences. The article is structured as followed: In section \ref{SEC::METHOD}
we describe our modified gravity simulations, explain how we create
surface mass density maps of haloes and introduce our computer vision based
image characterisation algorithm together with a simple classifier. We
demonstrate our ability to distinguish the different modified gravity models 
based on this approach in section \ref{SEC::RESULTS} and discuss these 
results in 
section \ref{SEC::DIS}. We conclude in section \ref{SEC::CONCL}
and give an outlook on future applications and improvements of the method.


\section{Methodology}
\label{SEC::METHOD}
In the following we describe the different steps necessary to use
computer vision based image characterisation to classify haloes into  
underlying structure formation models. Our main data set is a suite of 
simulations, carried out in eight
different models of gravity and using the same initial conditions. From each
simulation we choose the most massive systems and produce 2D maps of their
projected matter distribution. These maps are smoothed and converted into a
standard computer image format. A computer vision algorithm then extracts
almost 3000 features from the image into a feature vector.
All feature vectors per simulation class form a training set, which is used to
classify a test set based on a simple distance metric in image feature space.
\subsection{Numerical simulations}
\label{SEC::METHOD::SIMS}
Our suite of numerical simulations is described in \citet{Winther2015a} and 
has been used for scientific analyses in
\citet{Mead2016}, \citet{LHuillier2017} and \citet{vonBraun-Bates2017}.
All simulations
were carried out with a modified version \citep{Llinares2014} of the 
adaptive mesh refinement code
Ramses \citep{Teyssier2002} and implement four different models of gravity.
Common to all runs is the box size 250 Mpc/h, the number of $512^{3}$ particles
and the evolution of the cosmological background which is assumed to follow a
linear $\Lambda$CDM model with $(\Omega_{m}, \Omega_{\Lambda}, h_{100}) =
(0.267,0.733,0.704)$. All simulation runs were started with identical initial
conditions, a most crucial requirement for our study.

Since all the simulations we consider here have the same 
background evolution, all the differences between 
the modified gravity simulations and $\Lambda$CDM stem from the presence of a 
fifth-force in the former simulations. The modified gravity models also have 
what is known as a screening mechanism which is a way of dynamically 'hiding' 
the modifications of gravity in high density regions (relative to the cosmic 
mean). This allows these models to pass the stringent constraints from tests of 
gravity in the Solar System and at the same time giving rise to large 
modifications on cosmological scales. One can think of these models as 
effectively introducing a modified gravitational law of the form
$F = \frac{GMm}{r^2}(1 + C\cdot\epsilon\cdot e^{-r/\lambda})$.
Here $C$ describes the strength of the fifth-force relative to standard 
gravity, $\lambda$ is the interaction range of this force and $\epsilon$ is a 
screening factor. The screening factor is $1$ if the masses are small and/or 
are located in a low density environment. Otherwise the 
screening factor becomes $\epsilon \ll 1$. 
What constitutes small and large mass haloes in this context depends on the 
model parameters.

The first gravity model we consider is following standard Newtonian gravity 
($C=0$) and 
we will dub
this model as \textit{lcdm} in the following.
The second class are Hu-Sawicki $f(R)$ models \citep{Hu2007}, where the 
critical mass for a halo to be screened or not is $M_{\rm crit} \sim 
10^{13}(f_{R0}/10^{-6})^{1.5}M_{\odot}$. 
In this study we look into
two of these models, the first one with
$|f_{R0}|= 10^{-5}$ and another one with $|f_{R0}|= 10^{-6}$. We label the two
models as \textit{f5} and \textit{f6}, respectively.
The  coupling strength of these models is $C = \frac{1}{3}$ and they have an 
interaction range at the present time of $8$ Mpc$/h$ ($2$ Mpc$/h$) for 
\textit{f5} (\textit{f6}) which becomes smaller and 
smaller the further back in time we go. Hence, we expect stronger
deviations from \textit{lcdm} in the \textit{f5} model compared to \textit{f6}.
We also consider two different manifestations of Symmetron gravity 
\citep{Hinterbichler2010}, labelled \textit{symmA} and \textit{symmB},
with a coupling $C 
= \sqrt{1 - (a_{\rm SSB}/a)^3}$ for $a<a_{\rm SSB}$ and $C = 0$ otherwise. 
The interaction range is $\lambda=1$ Mpc$/h$. The difference between the models 
is the
scale factor of the symmetry breaking which is $a_{\mathrm{SSB}}=0.5$ for
\textit{symmA} and  $a_{\mathrm{SSB}}=0.33$ for \textit{symmB}. The 
modifications of gravity do not start to take effect 
before the scale factor is greater than $a = a_{\mathrm{SSB}}$ so
we expect larger effects for smaller values of $a_{\mathrm{SSB}}$.
The final class of models are the normal-branch
Dvali–Gabadadze–Porrati (DGP) gravity models \citep{Dvali2000}.
The DGP models have an infinite 
interaction range (just like for normal gravity) and $C \approx \frac{1}{3(1 + 
r_cH(a))}$. The relevant parameter here is the cross-over scale $r_{c}$
above which the 5-dimensional nature of the theory becomes relevant. Here we
study three such scales with $r_{c}H_{0}=0.5$,  $r_{c}H_{0}=1.2$ and
$r_{c}H_{0}=5.6$, labelled \textit{dgp05}, \textit{dgp12}  and  \textit{dgp56},
respectively. We expect the larger changes in comparison to the \textit{lcdm}
baseline model for smaller values of $r_{c}H_{0}$.
The final difference between the models is in how the screening, 
the $\epsilon$ factor, behaves. For the Symmetron and $f(R)$ models it is 
determined by the local value of the gravitational potential $\Phi$ and in the 
DGP models it is determined by the local matter density.
For the concrete implementation of the changes in the underlying equations of 
gravity the interested reader can refer to \citet{Llinares2014} 
and \citet{Winther2015a}. For a more thorough 
discussion about how these models work and modified gravity in general see
\citet{Clifton2012} and references therein.

For all eight simulation runs we have snapshots at three different redshifts $z
= 0.0$, $z=0.5$ and $z=1.0$. For each box and at all redshifts a halo
catalogue was produced by the Rockstar halo finder \citep{Behroozi2013}. From 
each
of the eight halo catalogues at three different redshifts we choose the 200
most massive objects, which are not a subhalo of a more massive parent halo. At
$z = 0.0$ this results in a typical mass range  $1.0-7.6\times 10^{14}
M_{\odot}/h$, at $z = 0.5$ the 200 systems lie between $0.7-4.6\times 10^{14}
M_{\odot}/h$ and for the highest redshift $z = 1.0$ between $0.4-3.4\times
10^{14} M_{\odot}/h$.

\subsection{Image creation}
\label{SEC::METHOD::IMAGES}
From the particle positions and masses within each simulation snapshot we
create surface mass density maps by centring the coordinate system on a
user-specified origin and by rotating the coordinate frame to match an, again
user-specified, line-of-sight. We then define a cube of given side length
around the current origin and with its z-axis aligned with the current
line-of-sight. The surface mass density map is created by diving the x-y face
of the cube, which is perpendicular to the line-of-sight, in $N_{x}\times
N_{y}$ cells and by projecting all particles in the cube that lie within the
cell while moving along the line-of-sight. From the surface mass density maps
we create 8-bit
TIFF\footnote{\href{http://www.fileformat.info/format/tiff/index.dir}{
http://www.fileformat.info/format/tiff/index.dir}} images by normalising the
surface mass density values of the map with respect to a $[0,255]$ interval.

\subsection{Training and test sets}
\label{SEC::METHOD::SETS}
We create images of the 200 most massive haloes in each of the eight gravity
models and for all three redshifts. We use a cube size of 10 Mpc/h and we
sample the surface mass density with $256\times 256$ cells. We will refer to
these cells as pixels from now on.
Since we can conveniently exploit the fact that we are working with projected
quantities, we use a total of 20 randomly chosen lines-of-sight per halo and
hence create for each gravity model and at all redshifts a training set class
of 4000 images. In order to test the effects of smoothing on the images we
produce six different realisations of each training set, the original
un-smoothed version and versions with a Gaussian smoothing with standard
deviations of 1, 2, 3, 4 and 5 pixel(s), respectively.

We also create test sets of images to check the performance of the classifier
later on. For these, we again use the 200 most massive haloes, for each gravity
model and at all redshifts, but along single a line-of-sight which differs from 
the ones used for the training sets. In order to include a 
simple noise model we apply Gaussian white noise to each cell of the surface
mass density map with signal to noise ratios (SNR) in each pixel of $\infty$ (no
noise applied), 20, 10, 5, 2 and 1 respectively. To all test set images, for
all models, redshifts and shot noise levels, we again produce six smoothed
realisations using the same Gaussian filters as for the training set.

In summary, we create a training set with eight classes, each class containing
4000 images. For each training set we create six versions with increasing levels
of smoothing applied. We create the same number of test set classes, with the
difference that they are based on only 200 images per class and that they have
different levels of noise applied. These images, split into a test and a
training set, containing eight classes each and available at three different
redshifts, are the starting point for the subsequent characterisation procedure.

\subsection{Characterisation}
\label{SEC::METHOD::CHAR}
We use the publicly available software
\texttt{WND-CHARM}\footnote{\href{https://github.com/wnd-charm/wnd-charm}{
https://github.com/wnd-charm/wnd-charm}} to characterise the halo images. This
software was originally designed for medical and biological applications,
especially the classification of objects seen under a microscope
\citep{Shamir2008}. Many other such algorithms or complete software packages
exist \citep[e.g.][for a variety of different
approaches]{Bengtsson2003,Heller2006,Yavlinsky2006}, but a unique feature of
\texttt{WND-CHARM}  is the fact that it derives a large amount of image
descriptors of very different kinds. It is thereby not limited to a specific
set of image features, such as e.g. image textures, and hence renders it ideal
for our aim in this pilot study: a fully agnostic view on the classification
problem using computer vision with very little a-priori knowledge on what the
most descriptive image features will be.

The monochrome, sometimes called grey-scale,  $256\times 256$ pixel 8-bit TIFF
images are converted by \texttt{WND-CHARM} into an image feature vector. While
doing so it does not only operate on the raw pixel intensity data, but also on
several transformations of the image including its Fourier (F), Chebyshev (C)
and wavelet (W) transformation. The latter is produced using a one level
filter pass and a symlet of order 5 \citep{Orlov2008}. It also considers an
Edge (E) transformation, a standard transformation in digital image processing,
which is carried out by approximating the image gradient with the Prewitt
operator \citep{Prewitt1970a}. Finally, also transformations of transformations
are analysed including the Chebyshev transformation of the image's Fourier
transform, C(F), as well as W(F), F(W), F(C), C(W), F(E) and W(E).

A large set of features, in total 2919, is extracted from the images and its
transforms. We list them in table \ref{TAB::METHOD::FEATURES}, where we also
show from which of the numerous image transformations a feature is derived
from. The total feature vector can be divided into
feature families, each of which with its own set of feature classes. We will
describe this hierarchy in the following and strive on the description of each
individual feature. For more information on specific image features,  we refer 
the interested reader to the
literature provided by the references in the following.

The first family of features is pixel statistics. The simplest class of such
features that \texttt{WND-CHARM} derives are simple pixel intensity statistics
and consists of mean, median, standard deviation, minimum and maximum of the
image's intensity values. This provides five features per image and image
transform. Slightly more complex are multi-scale histograms of the intensity
distribution using three, five, seven or nine bins, respectively. The values for
each bin add 24 features per image and transform. The first class of features
which contains information on the spatial correlation of intensity values are
the combined moments. Mean, standard deviation, skewness and kurtosis are
calculated from all intensity values that fall within a stripe along the x-axis
through the image centre with a width that is half the total image height. This
stripe is then rotated by 45, 90, and 135 degrees around the centre, creating
another three sets of first moments and sampled into three-bin histograms. A
total of 48 features in this class is the result. More abstract is the Gini
coefficient of the image, which was originally defined for economical studies,
but which is now often used in astrophysical applications
\citep[e.g.][]{Florian2016}. The full definition can be found in
\citep{Abraham2003}, in short, the Gini coefficient describes with a single
number the level of discrepancy of an image's intensity value distribution from 
a
perfectly equal intensity distribution.

The second family of feature classes derives from polynomial decompositions of
the image intensities and some of its transforms. The first class of these
decompositions are based on the Chebyshev coefficients from an order $N=20$
Chebyshev transform. The values of the coefficients are used to fill a 32-bin
histogram, providing the 32 image features of this class. Another set of 32
features comes from the coefficient histogram of an order $N=23$
Chebyshev-Fourier transform \citep{Orlov2006} of the image. After a Zernike
decomposition of the 2D image \citep{Teague1980}, the 72 first Zernike
coefficients create another class of features. Finally, the radon
transformations \citep{Radon1917} along lines with inclination angle 0, 45, 90
and 135 degrees are binned into 3-bin histograms which provide another twelve
features per image and image transformation.

Image textures, the third feature family, describe the spatial correlation of 
intensity
values. \texttt{WND-CHARM} applies Gabor filters \citep[e.g.][]{Fogel1989} for
seven frequencies and using a Gaussian harmonic function as a kernel following
\citep{Grigorescu2002}. This provides the seven Gabor texture features
which are defined as the area, occupied by the Gabor-transformed images
\citep{Orlov2008}. Tamura textures describe contrast, coarseness and
directionality of an image. We refer the interested reader to
\citet{Tamura1978} for a full definition of these properties. Contrast and
directionality are two of the Tamura features used in this analysis, coarseness
sum and a 3-bin coarseness histogram complete the set of six features in this
class. Another, quite specific, feature class are Haralick textures, which are
described in \citet{Haralick1973}. As many texture features, they are
calculated from the  spatial grey-level dependence matrix (SGLDM) of the image
which encodes how many times per image specific intensity value pairs are 
located at a certain distance and along a certain direction. Haralick
textures are the fundamental statistical properties of this matrix and contain
important features used in digital image processing such as the angular second
moment (ASM), the image contrast, correlation and entropy. The 28 properties of
the SGLDM defined in the appendix of \citet{Haralick1973} are the Haralick
features used by \texttt{WND-CHARM} and in this study. In addition to image
textures based on the  SGLDM, \citet{Wu1992} proposed to characterise the
roughness of an image by using a fractional Brownian motion model on multiple
scales. The 20 image features coming from the fractal model class implemented
in \texttt{WND-CHARM} complete the texture family of features

Probably more related to the human perception of images is the last family of
features which works on the raw pixel data only and tries to detect distinct
objects in the image. The edge feature class applies a Prewitt operator
\citep{Prewitt1970a} to approximate the image gradient. The mean, median,
variance and 8-bin histogram of both magnitude and directionality of the
gradient provide 22 image features. They are complemented by the total number
of edge pixels, their direction homogeneity and a 4-bin histogram of all
possible differences between the directionality histogram bins, that were
mentioned earlier. The final class of object features are Otsu features. They
are calculated after applying a global Otsu threshold \citep{Otsu1979} to the
image, thereby converting the image to binary. Basic statistics of all
eight-connected (connected to edge or corner) objects in this mask are then
collected including, their abundance, Euler number and centroid (both
coordinates). In addition, minimum, maximum, mean, median variance and a 10-bin
histogram are calculated for area and distance to the
centroid of all objects. In total 34 features. The same procedure is applied to 
the inverse
Otsu binary image, adding a final set of 34 features.

We refer the reader to table \ref{TAB::METHOD::FEATURES} for an overview of all
features and to which group of image transforms they are applied to.

\begin{table*}
\caption{Image features used in this analysis.}
\label{TAB::METHOD::FEATURES}
\begin{tabular}{cccccc}
\hline
Family & Class & Features & Input & Reference\\
\hline
\hline
Pixel statistics& Combined moments & 48 & raw, F, W, C, C(F), W(F) & -- \\
&&&F(W), F(C), C(W), E, F(E), W(E) & \\
& Gini coefficient & 1 & raw, F, W, C, C(F), W(F) & \citet{Abraham2003} \\
&&&F(W), F(C), C(W), E, F(E), W(E) & \\
& Multiscale histograms & 24 & raw, F, W, C, C(F), W(F) & -- \\
&&&F(W), F(C), C(W), E, F(E), W(E) & \\
& Pixel intensity statistics & 5 & raw, F, W, C, C(F), W(F) & -- \\
&&&F(W), F(C), C(W), E, F(E), W(E) & \\
\hline
Polynomial decomposition& Chebyshev coefficients & 32 & raw, F, W, C, F(W), E,
F(E), W(E) &--\\
& Chebyshev-Fourier coefficients & 32 & raw, F, W, C, F(W), E, F(E), W(E) &
\citet{Orlov2006}\\
& Radon coefficients & 12 & raw, F, W, C, C(F), W(F) & \citet{Radon1917}\\
&&&F(W), F(C), C(W), E, F(E), W(E) & \\
& Zernike coefficients & 72 & raw, F, W, C, F(W), E, F(E), W(E)&
\citet{Teague1980}) \\
\hline
Textures& Fractal analysis & 20 & raw, F, W, C, C(F), W(F) & \citet{Wu1992} \\
&&&F(W), F(C), C(W), E, F(E), W(E) & \\
& Gabor & 7 & raw & \citet{Fogel1989} \\
& Haralick & 28 & raw, F, W, C, C(F), W(F)&\citet{Haralick1973}\\
&&&F(W), F(C), C(W), E, F(E), W(E) & \\
& Tamura & 6 & raw raw, F, W, C, C(F), W(F)& \citet{Tamura1978} \\
&&&F(W), F(C), C(W), E, F(E), W(E) & \\
\hline
Objects& Edge features & 28 & raw & \citet{Prewitt1970a} \\
& Otsu object features & 34 & raw & \citet{Otsu1979}\\
& Inverse Otsu object features & 34 & raw & \citet{Otsu1979}\\
\end{tabular}
\end{table*}

\subsection{Classification}
\label{SEC::METHOD::CLASS}
The computer vision characterisation algorithm produces a feature vector $F$
for each image, from both training and test sets. The number of features in
this vector shall be $|F|=M$. One can now derive a simple classification scheme
based on a metric for the distance between a test image to the training set
classes in the M-dimensional feature space or a subset of it. The total number
of classes in the training set shall be $N$. Not every feature is equally
informative in discriminating one training set class from another. This is why
we assign weights $w_{f}$ to each single feature $f$ following a Fisher 
discriminant
criterion \citep[e.g.][]{Bishop2006a}
\begin{equation}
w_{f} =
\frac{\sum\limits_{c=1}^{N}\left(\left<F_{f}\right>-\left<F_{f}^{c}
\right>\right)^{2}}{\sum\limits_{
c=1}^{N}(\sigma_{f}^{c})^{2}}\frac{N}{N-1},
\label{EQU::METHOD::WEIGHT}
\end{equation}
where $\left<F_{f}\right>$ is the mean of all values of the feature $f$ among 
all images
in the training set and $\left<F_{f}^{c}\right>$ is the mean of all values of a 
specific
feature $f$ within a single class $c$ of the training set and
$(\sigma_{f}^{c})^2$ is its variance within that class. In this, all variances
are calculated after the feature values are normalised to fall into the
interval $[0,1]$.

Once all feature weights are calculated from the training set, we can calculate
the feature space distance of a single test image feature vector $F$ from a
training set class $c$. One example of such a distance metric which is readily
implemented in \texttt{WND-CHARM} is the weighted nearest neighbour (WNN)
distance
\begin{equation}
d_{\mathrm{WNN}}^{c} = \min\limits_{T\in T_{c}} \sum\limits_{f=1}^{M}
w_{f}\left( F_{f} -T_{f} \right)^{2},
\label{EQU::METHOD::WNN}
\end{equation}
where $T$ is a feature vector from the training set, $T^{c}$ is the set of all
feature vectors in the training set that belong to class $c$ and $w_{f}$ are
the feature weights as defined by equation \ref{EQU::METHOD::WEIGHT}. Another
feature distance metric, which is also readily implemented in
\texttt{WND-CHARM}, is the weighted neighbour distance \citep{Orlov2008}
\begin{equation}
d_{\mathrm{WND}}^{c} = \frac{\sum\limits_{T\in
T_{c}}\left[\sum\limits_{f=1}^{M} w_{f}\left( F_{f} -T_{f}
\right)^{2}\right]^{p}}{|T^{c}|}
\label{EQU::METHOD::WND}
\end{equation}
where $|T^{c}|$ is the number of training images in class $c$ and $p$ is a free
parameter. The main difference between the WNN and WND distance metric is that
WNN defines the distance between a feature vector $F$ and a class $T^{c}$ by
taking into account only the smallest of all distances (or generally k-smallest
distances) between $F$ and all $T\in T^{c}$. In contrast the WND metric takes
into account all distances to the images of the training set of class $c$, but
suppresses the larger contributions by means of the exponent $p$.
It was shown in \citet{Orlov2008} that the classification accuracy is not
particularly sensitive to $p$ as long as it is smaller than $-4$. In the
\texttt{WND-CHARM} package it is fixed to $-5$. \citet{Orlov2008} also showed
that the WND metric gives slightly better results than WNN in biomedical image
classification problems.

Once we decided on a distance metric we can define the similarity of a given
test image described by its feature vector $F$ to the classes $c$ defining the
training set as
\begin{equation}
S_{F}^{c} = \left(d_{F}^{c}\sum\limits_{i=1}^{N}(d_{F}^{i})^{-1} \right)^{-1}
\label{EQ::METHOD::SIM}
\end{equation}
This conveniently assigns for each test image--class pair a similarity
$S_{F}^{c} \in [0,1]$. The classification is then performed by assigning the
class with the highest similarity to the test image.

At this point it has to be mentioned that other classification schemes are
possible, e.g. based on machine learning (including support vector 
machines, neural networks or Gaussian processes) and potentially implemented as 
a Bayesian scheme \citep[e.g.][]{Bishop2006a}. We chose a simple
feature distance based scheme since we wanted to retain a maximum number of
usable features while achieving acceptable runtimes in this exploratory study.
Other classification schemes shall be explored in future work, the general
methodology is not limited to our rather simple current approach.

A more detailed description of \texttt{WND-CHARM}, the software used for image
characterisation and classification in this work is given in \citet{Orlov2008}.
A specific description of the software package and its usage is given by
\citet{Shamir2008} and examples for its applications in bio medical and
astronomical classification problems can be found in \citet{Shamir2010} and
\citet{Schutter2015a}.


\section{Results}
\label{SEC::RESULTS}
The methodology described in sections \ref{SEC::METHOD::CHAR} and
\ref{SEC::METHOD::CLASS} is applied to the simulated data described in section
\ref{SEC::METHOD::SIMS}. Each of the eight classes in the training set contains
4000 halo images and comes with a set of 200 halo images for testing the
classifier. The test sets are available for six different signal to noise
levels. Every class has six different smoothing lengths applied to it and is
analysed at three different redshifts.

In this section we first identify the most severe degeneracies between the
models and reduce the number of classes to four. We then report on the
classification success rates as a function of external parameters such as
signal to noise in the images and applied smoothing scale and as a function of
physical parameters such as redshift or halo mass.

\subsection{Model degeneracies}
Some of the modified gravity models are expected to give very similar results
given the choice of their parameters. We hence identify the most severe
degeneracies in order to discard models from this initial study which are
almost impossible to distinguish from one another based on image morphology 
characterisation only. As a first step we therefore run a series of 
classifications with only
two model classes and cycle through all possible combinations of model pairs.
The success rates of these classification runs, so the ratio between the
correctly classified haloes in a certain class and the total number of 
tested haloes is reported in table \ref{TAB::RES::MODEL1ON1}. In
this first set of classifications we choose the noise-free image samples and
apply no smoothing. To perform the classification we use the WND distance
metric on the full feature vector containing 2919 image features. This set up
is fiducial for classification runs and used if no further comments
are made.

\begin{table*}
\caption{Classification success rates for all possible direct model pair
comparisons.}
\label{TAB::RES::MODEL1ON1}
\begin{tabular}{l|cccccccc|}
\hline
&lcdm&f5&f6&symmA&symmB&dgp05&dgp12&dgp56\\
\hline
lcdm&*&0.83&\textbf{0.70}&\textbf{0.66}&0.84&0.79&0.78&0.80\\
f5&*&*&0.8&0.76&0.78&0.83&0.82&0.82\\
f6&*&*&*&\textbf{0.53}&0.77&0.79&0.76&0.79\\
symmA&*&*&*&*&0.79&0.79&0.77&0.80\\
symmB&*&*&*&*&*&0.87&0.82&0.84\\
dgp05&*&*&*&*&*&*&\textbf{0.61}&\textbf{0.72}\\
dgp12&*&*&*&*&*&*&*&\textbf{0.62}\\
\end{tabular}
\end{table*}

In this case, the expected success rate for a classification by random pick is
$0.5$ and the table highlights in bold font the rates for all model combinations
which are smaller then $0.75$. As expected, the \textit{f6} and \textit{symmA}
models differ only mildly from \textit{lcdm}. The degeneracy between
\textit{symmA} and \textit{f6} is even stronger. Interestingly, the three DGP
models are particularly distinct from e.g. \textit{lcdm}, but they are quite
degenerate among themselves. As a main result of this analysis we drop in the
following the \textit{f6}, \textit{symmA}, \textit{dgp05} and \textit{dgp12}
models. From now on we will only focus on a single model per modified gravity
class namely \textit{lcdm}, \textit{f5}, \textit{symmB} and \textit{dgp56}.
The choice of \textit{lcdm} is clear since it is the baseline
cosmology. \textit{f5} and \textit{symmB} are chosen since they eliminate the
worst degeneracies to \textit{lcdm} within their gravity model class. The
choice of \textit{dgp56} is more arbitrary but is based on the fact that all
DGP models seem similarly degenerate to the other gravity models and
\textit{dgp56} should give the smallest differences compared to \textit{lcdm}.
We hence picked it as a worst case scenario. We will, however, revisit
this choice later in the analysis. For completeness, we show a number of key 
analyses for the full set of eight models in appendix \ref{SEC::APP}.

\subsection{General classification success rates}
\label{SEC::RES::GENERAL}
The main classification result for the remaining four models is shown in table
\ref{TAB::RES::MAIN}. We report there the fiducial classification run at $z=0$,
with no added noise and with no smoothing applied. Furthermore, the full
feature vector with 2919 entries was used to calculate the similarities. Each
main row of the table represents a class of test sets. The first line in each
of these rows shows the number of images which were classified as a member of
the respective training set class indicated by the column. The second line in
each model row shows the average similarity value (see equation
\ref{EQ::METHOD::SIM}) for all test images of the current test class to every
training set class. The last line in the model
row is just this similarity value normalised to the similarity of the correct
test image class. Finally, also the overall success rate per model is shown in
the last column of each model row and at the very bottom of the table we report
the total success rate of the classification run.

The overall success rate is 61\%, far better than the expected result of 25\%
for a classification by random pick. The results for the individual models are
quite different. 122 out of 200 \textit{lcdm} haloes were classified correctly
and by looking at the average similarities, the model appears quite distinct
from \textit{dgp56}; it is mildly degenerate with \textit{f5} with 26
mis-classifications and more severely degenerate with \textit{symmB} with 42
mis-classifications out of 200 haloes. \textit{f5} appears to be the most
difficult model to classify. Although its success rate of 46\% still clearly
favours the correct answer, there is a strong degeneracy with \textit{symmB}
and a mild degeneracy with \textit{lcdm}. Again, there are almost no
mis-classifications of \textit{f5} images as \textit{dgp56}. The best result is
indeed seen for the \textit{symmB} classification, with 142 our of 200 haloes
or 71\%. A small and equal trend to mis-identify \textit{symmB} haloes as 
\textit{lcdm}
or \textit{f5}, respectively is apparent. Also \textit{dgp56} is classified
solidly with a 65\% success rate. Most interestingly there is a pronounced
degeneracy with \textit{lcdm}, something that is not seen vice-versa.

\begin{table}
\caption{Classification matrix for four models using the full feature vector at
no added noise or smoothing.}
\label{TAB::RES::MAIN}
\begin{tabular}{l|ccccc|}
\hline
&lcdm&f5&symmB&dgp56&success rate\\
\hline
\hline
       &122&26&42&10\\
lcdm &0.3698&0.2535&0.2422&0.1345&0.61\\
	 &1.00&0.69&0.66&0.36&\\
\hline
 	&34&92&66&8&\\
f5  &0.2467&0.3579&0.2904&0.1049&0.46\\
	&0.69&1.00&0.81&0.29&\\
\hline
		&27&25&142&6&\\
symmB	&0.2374&0.2801&0.3829&0.0996&0.71\\
		&0.62&0.73&1.00&0.26&\\
\hline
 		&40&12&19&129\\
dgp56   &0.1709&0.1414&0.1361&0.5516&0.65\\
		&0.31&0.26&0.25&1.00&\\
\hline
\hline
all &&&&&0.61
\end{tabular}
\end{table}

\subsection{Classification success rate dependence on external parameters}
\label{SEC::RES::EXTERNAL}
\begin{figure}
\includegraphics[width=.5\textwidth]{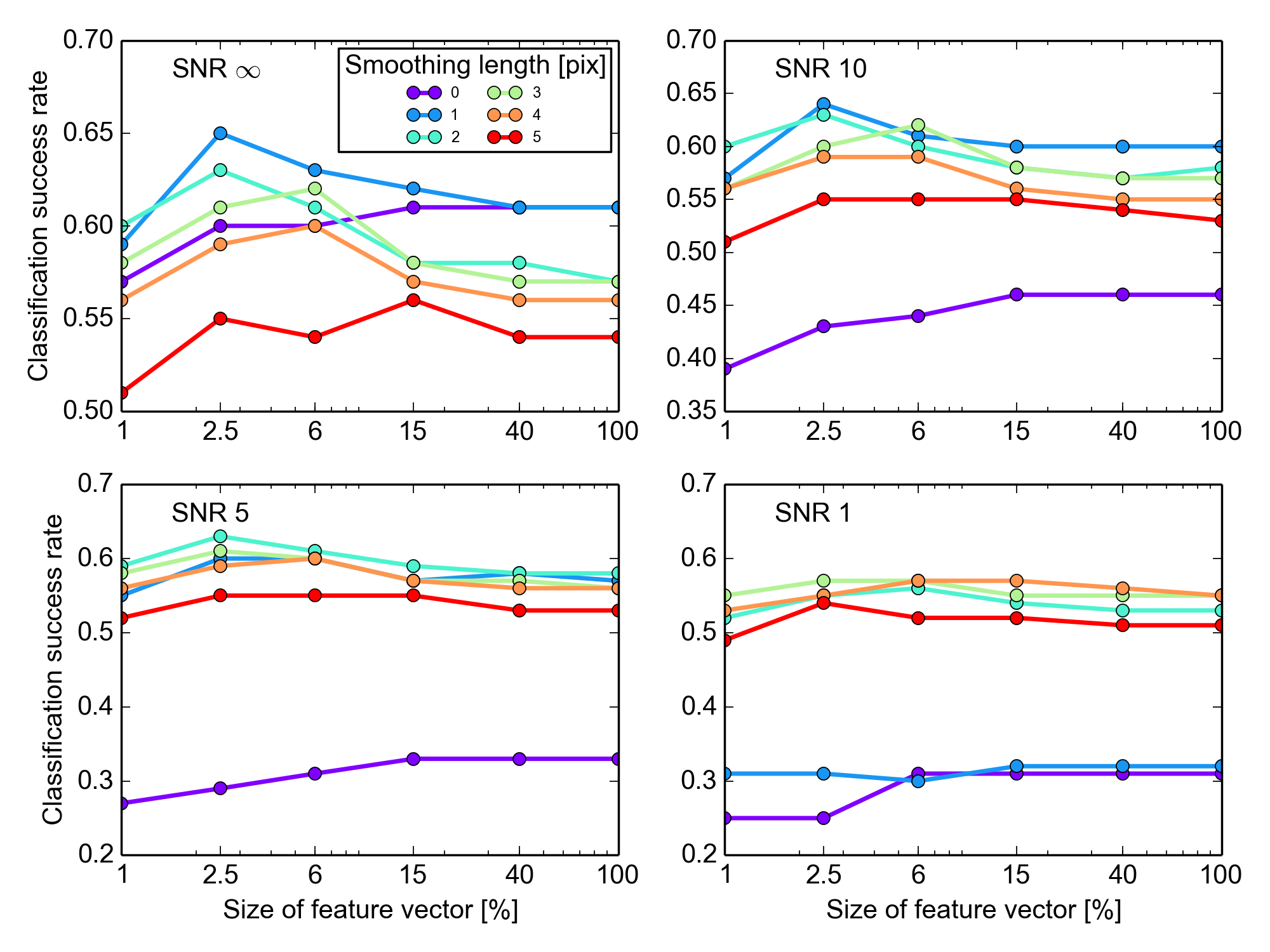}
\caption{The overall classification success rate as a function of several
external parameters. The four panels show classification runs with different
levels of noise applied to the test images, as indicated in the top left corner
of each panel. The colour-coded lines show different levels of smoothing applied
to both training and test images and the x-axis in each panel refers to the
percentage of fisher-weight ranked image features that were used for the
classification, where 100\% is 2919 features.}
\label{FIG::RES::SML_PARAM_SNR}
\end{figure}

In contrast to the fiducial classification, figure
\ref{FIG::RES::SML_PARAM_SNR} shows  the overall classification success rate
while varying some external parameters. These parameters include the level of
white noise that was added to the test images, the level of smoothing that was
applied to the training and test images and the number of parameters that were
used to perform the classification. Figure \ref{FIG::RES::SML_PARAM_SNR} shows
a number of expected results. Firstly, the classification success rate drops
quickly in the presence of increasing noise levels. From the familiar
$\sim$60\% in the absence of noise, down to less than 30\% for very noisy
images with a signal to noise ratio of $1$. However, this trend can readily be
fixed with the application of mild levels of smoothing. A rather small
smoothing scale of one pixel is enough to bring the classification rate back to
about 60\%. In the extreme case of a signal to noise ratio of $1$, smoothing on
a scale of three pixels is needed to achieve this. While smoothing is crucial
in the presence of noise, it only mildly affects the classification rate in the
absence of noise. Even for a rather extreme smoothing scale of five pixels, the
success rate drops by only 7\%. Most interestingly, the best overall
classification result is not achieved for the fiducial case, but for a
smoothing scale of one pixel, while using only the top 2.5\% of the elements in
the  fisher-weight ranked feature vector, so the 73 most discriminating
features.

\begin{figure}
\includegraphics[width=.5\textwidth]{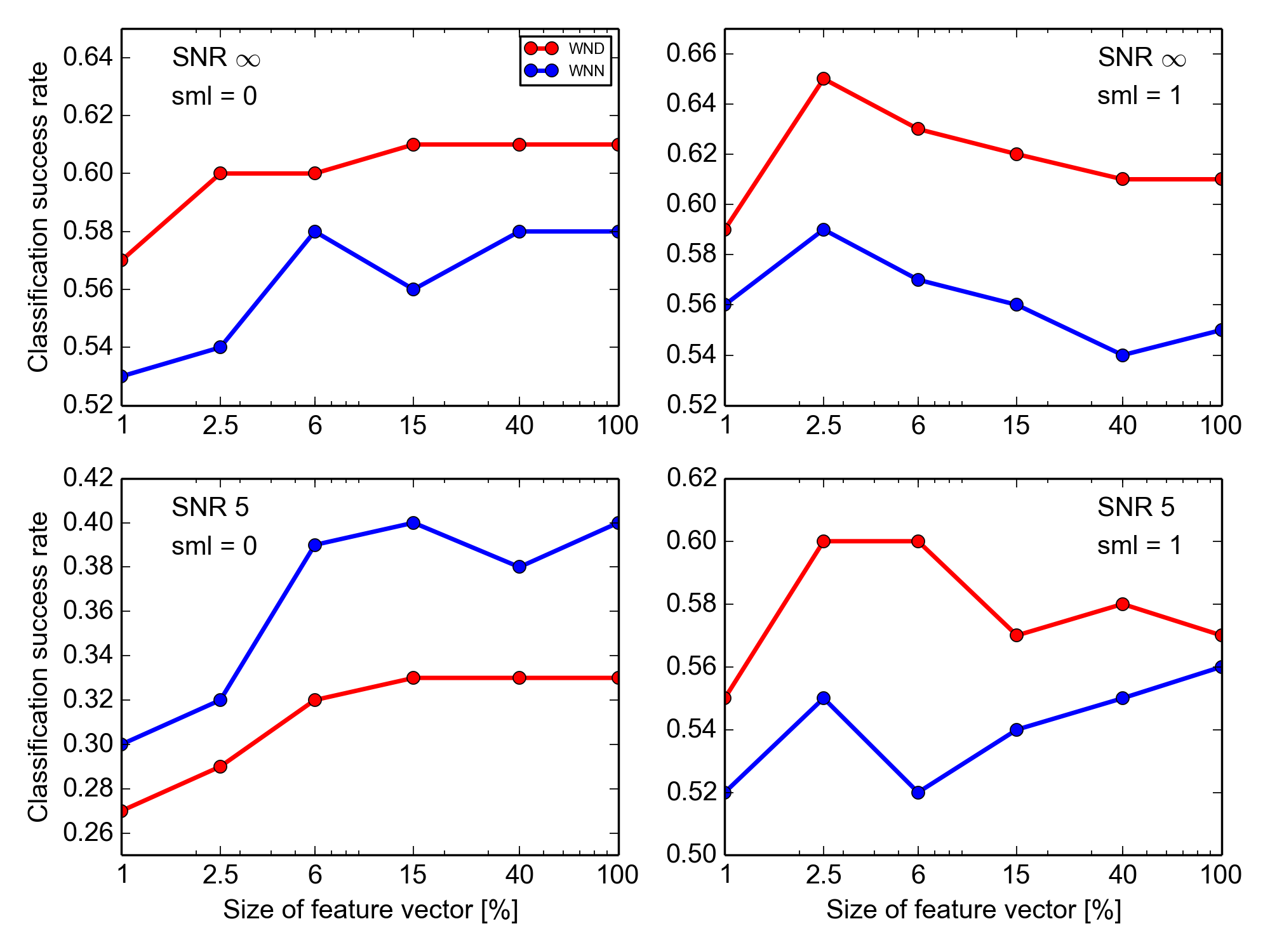}
\caption{The overall classification success rate as a function of distance
metric. Similar to figure \ref{FIG::RES::SML_PARAM_SNR}, the four panels show
different levels of noise in the test images, but in this plot they also encode
a fixed smoothing scale that was applied to both, test and training set images.
The size of the feature vector is given as percentage relative to the full 2919
features by the x-axis. The WND  distance metric (equation
\ref{EQU::METHOD::WND}) is shown in red, the WNN distance metric (equation
\ref{EQU::METHOD::WNN}) is shown in blue.}
\label{FIG::RES::METRIC}
\end{figure}

Figure \ref{FIG::RES::METRIC} is similar to figure
\ref{FIG::RES::SML_PARAM_SNR} but shows the overall classification success rate
for both, the WND and WNN distance metric. We confirm the trend reported by
\citet{Orlov2008}, that the WND is generally superior to WNN. The only
exception are cases of low signal to noise in the images and in the absence of
smoothing (bottom-left panel of figure \ref{FIG::RES::METRIC}). However, this
configuration delivers generally bad results as discussed earlier and
highlighted in figure \ref{FIG::RES::SML_PARAM_SNR}. It seems that in this case
the WND strategy of considering the distances to all training images of a class
can indeed introduce significant biases in the presence of noise. We will
investigate this behaviour in more detail in the discussion of the following
section.

\subsection{Classification success rate dependence on physical parameters}
\label{SEC::RES::PHYS}
In order to study the classification success rate as a function of physical
properties we further divide each of the four model classes into four mass bins
of equal object number. The first bin contains the 50 most massive systems in
the sample, the second mass bin the $50^{\mathrm{th}}$ to $100^{\mathrm{th}}$
most massive system and so on. The overall classification results and the
results for each individual model are reported in table
\ref{TAB::RES::MASS_BINS}. As can be seen there, the classification success can
indeed increase by up to 10\% when compared to the fiducial single mass bin
run. The only exception is the first, most massive mass bin, where we have a
wide mixture of masses due to the steep drop of the mass function at the
high-mass end. For the other mass bins, where the mass separation is more
effective, the increase is more pronounced.

\begin{table}
\caption{Classification success rate in bins of halo mass}
\label{TAB::RES::MASS_BINS}
\begin{tabular}{lccccc}
\hline
mass bin&1&2&3&4&all\\
\hline\hline
lcdm&0.64&0.74&0.64&0.72&0.61\\
f5&0.52&0.60&0.50&0.54&0.46\\
symmB&0.66&0.74&0.74&0.66&0.71\\
dgp56&0.68&0.74&0.90&0.80&0.65\\
\hline
all&0.63&0.71&0.70&0.68&0.61\\
\end{tabular}
\end{table}

Table \ref{TAB::RES::REDSHIFT_BINS} shows  the overall and individual model
classification success rate in the three available redshift bins. We see a
continuous decrease with increasing redshift, related to the fact that at high
redshift the distinguishing effects of each model had less time to act on the
morphology. In the course of this redshift dependence study and as 
an
additional exercise we attempted a classification by redshift. We take all
images of the \textit{lcdm} model and group them into redshift classes. We then
use these classes and the familiar characterisation and classification scheme
to determine the redshift for each of the respective test images. The results
are reported in table \ref{TAB::RES::Z}, which follows the same layout as table
\ref{TAB::RES::MAIN}. The success rate at low redshift is almost 100\% and we
find a good 80\% for our highest redshift class. The intermediate redshift
class gives the worst result since it is bracketed by two instead of only one
other model and is hence more prone to degeneracies. Still, a  74\%
success rate shows the ability to sort objects by redshift based on their image
morphology only.

\begin{table}
\caption{Classification success rate at different redshifts}
\label{TAB::RES::REDSHIFT_BINS}
\begin{tabular}{lccc}
\hline
z&0.0&0.5&1.0\\
\hline \hline
lcdm&0.61&0.51&0.46\\
f5&0.46&0.58&0.52\\
symmB&0.71&0.55&0.39\\
dgp56&0.65&0.64&0.55\\
\hline
total&0.61&0.57&0.48\\
\end{tabular}
\end{table}

\begin{table}
\caption{Classification matrix for redshift classes.}
\label{TAB::RES::Z}
\begin{tabular}{l|cccc|}
\hline
&z00&z05&z10&success rate\\
\hline
       &193&7&0&\\
z00 &0.8638&0.1268&0.0094&0.97\\
	 &1.00&0.15&0.01&\\
\hline
       &26&148&26&\\
z05 &0.1558&0.6217&0.2225&0.74\\
	 &0.25&0.35&1.00&\\
\hline
       &1&40&159&\\
z10 &0.0120&0.2554&0.7326&0.80\\
	 &0.02&0.35&1.00&\\
\hline
\hline
all &&&&0.83
\end{tabular}
\end{table}


\section{Discussion}
\label{SEC::DIS}
We now revisit and elaborate on the more interesting results of section
\ref{SEC::RESULTS}.

\subsection{Optimal parameter choices}
\label{SEC::DIS::OPTIMAL}
As figure \ref{FIG::RES::SML_PARAM_SNR} implies, there is a benefit of applying
a mild one pixel smoothing to images, even if they are noise free, and then
limit the number of features to be used for the classification to about 2\% of
the total feature vector. With the optimal choice of smoothing and feature 
vector length, we obtain the classification result that is shown in table
\ref{TAB::DIS::OPTIMAL}. The overall classification success rate raises
slightly by 4\% to 65\%. The largest individual model increases in the success
rate are seen for \textit{dgp56}, from 65\% to 74\% and for \textit{f5}, from
46\% to 52\%. It seems that the combination of smoothing and feature reduction
removes a number of degeneracies to other models, mostly to \textit{symmB} for
\textit{f5} and to \textit{lcdm} for \textit{dgp56}.

 \begin{table}
\caption{Classification matrices in the absence of noise with optimal choice of
smoothing scale (one pixel) and feature vector size ($\sim$70).}
\label{TAB::DIS::OPTIMAL}
\begin{tabular}{l|ccccc|}
\hline
&lcdm&f5&symmB&dgp56&success rate\\
\hline
       &129&27&31&13\\
lcdm &0.4704&0.2065&0.1935&0.1295&0.65\\
	 &1.00&0.44&0.41&0.28&\\
\hline
 	&39&104&53&4&\\
f5  &0.2167&0.4287&0.2672&0.0874&0.52\\
	&0.51&1.00&0.62&0.20&\\
\hline
		&32&22&140&6&\\
symmB	&0.1701&0.2497&0.5028&0.0773&0.70\\
		&0.34&0.50&1.00&0.15&\\
\hline
 		&26&9&17&148\\
dgp56   &0.1492&0.1010&0.0979&0.6519&0.74\\
		&0.23&0.15&0.15&1.00&\\
\hline
\hline
all &&&&&0.65
\end{tabular}
\end{table}

\subsection{Response to noise}

The WND distance metric turned out to be the best choice in most cases. The
only exception that shows up in figure \ref{FIG::RES::METRIC} is the low
signal to noise case without subsequent smoothing. To study this case in more
detail we show the success rate for individual classes in table
\ref{TAB::DIS::METRIC}. Most interestingly, the WND metric fails completely to
classify successfully the \textit{f5} model in the absence of smoothing, but
excels over WNN in the case of \textit{dgp56}. It appears that the strategy of
WND to take into account all training images of a class, even while suppressing
the contribution of the more distant samples in feature space, introduces a
severe bias in the treatment of \textit{f5} images. When looking into the
results in more detail, it shows that 128 out of the 200 \textit{f5} images get
classified as \textit{dgp56} in this case.
This is a first hint of irregularities with the \textit{dgp56} model in the
presence of noise and in the absence of smoothing, which we shall investigate
further.

\begin{table}
\caption{Classification success rates for different settings of external
parameters and depending on the distance metric used for classification.}
\label{TAB::DIS::METRIC}
\begin{tabular}{ccc}
\hline
&WND&WNN\\
\hline
\hline
SNR = 5, no smoothing &&\\
\hline
lcdm &0.18&\textbf{0.41}\\
f5 &0.0&\textbf{0.15}\\
symmB &0.27&0.27\\
dgp56 &\textbf{0.88}&0.77\\
\hline
total &0.33&\textbf{0.40}\\
\hline
SNR = 5, 1 pixel smoothing &&\\
\hline
lcdm &\textbf{0.61}&0.50\\
f5 &0.40&0.40\\
symmB &\textbf{0.70}&0.59\\
dgp56 &\textbf{0.71}&0.70\\
\hline
total &\textbf{0.60}&0.54 \\
\end{tabular}
\end{table}

\begin{figure}
\includegraphics[width=.5\textwidth]{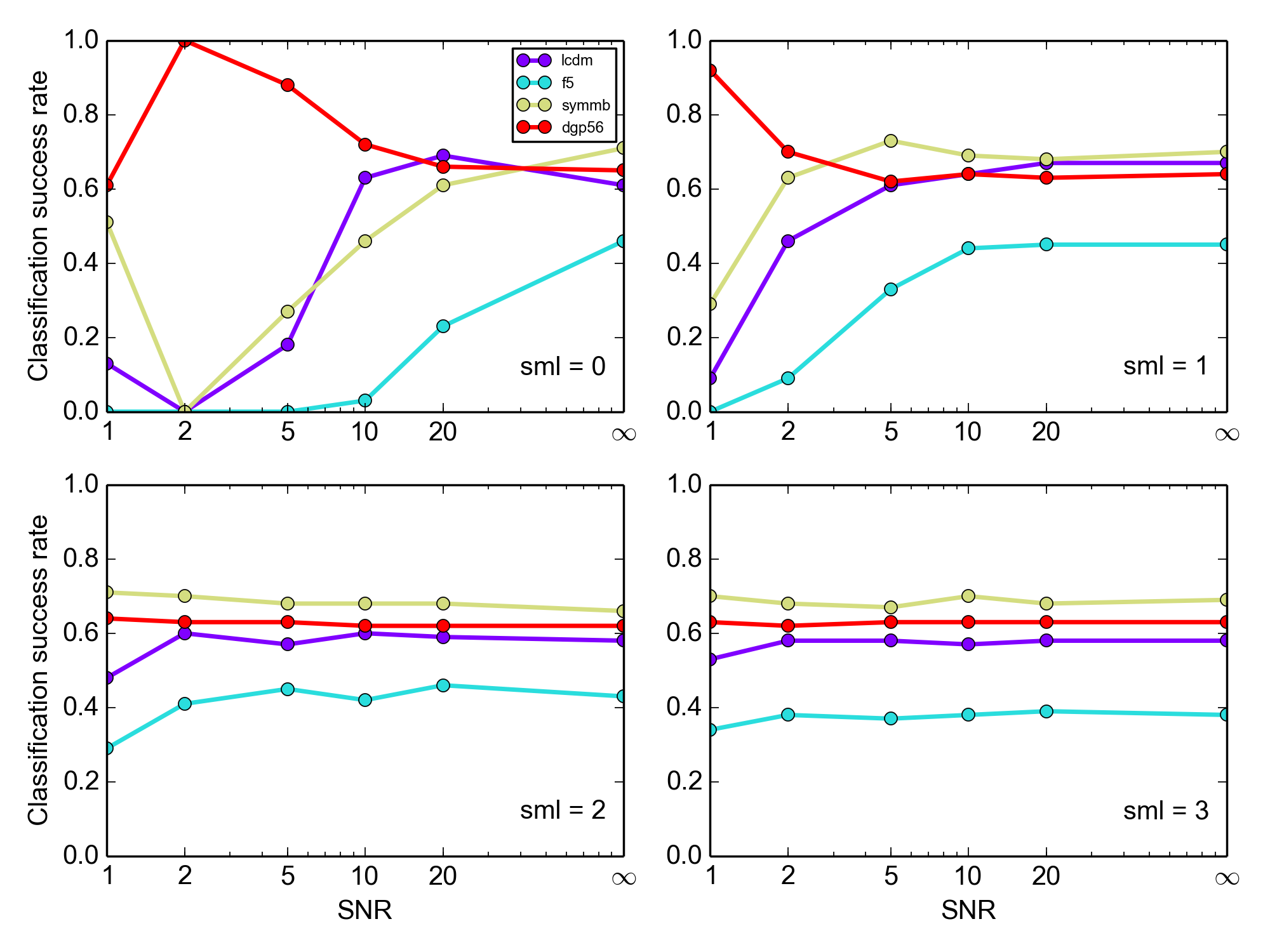}
\caption{Classification success rate for individual models as a function of
signal to noise of the test images. The four different panels refer to four
different smoothing scales applied to training and test data, indicated by the
label in the bottom-right corner of each panel.}
\label{FIG::DIS::MODEL_SUCCESS}
\end{figure}

\begin{figure}
\includegraphics[width=.5\textwidth]{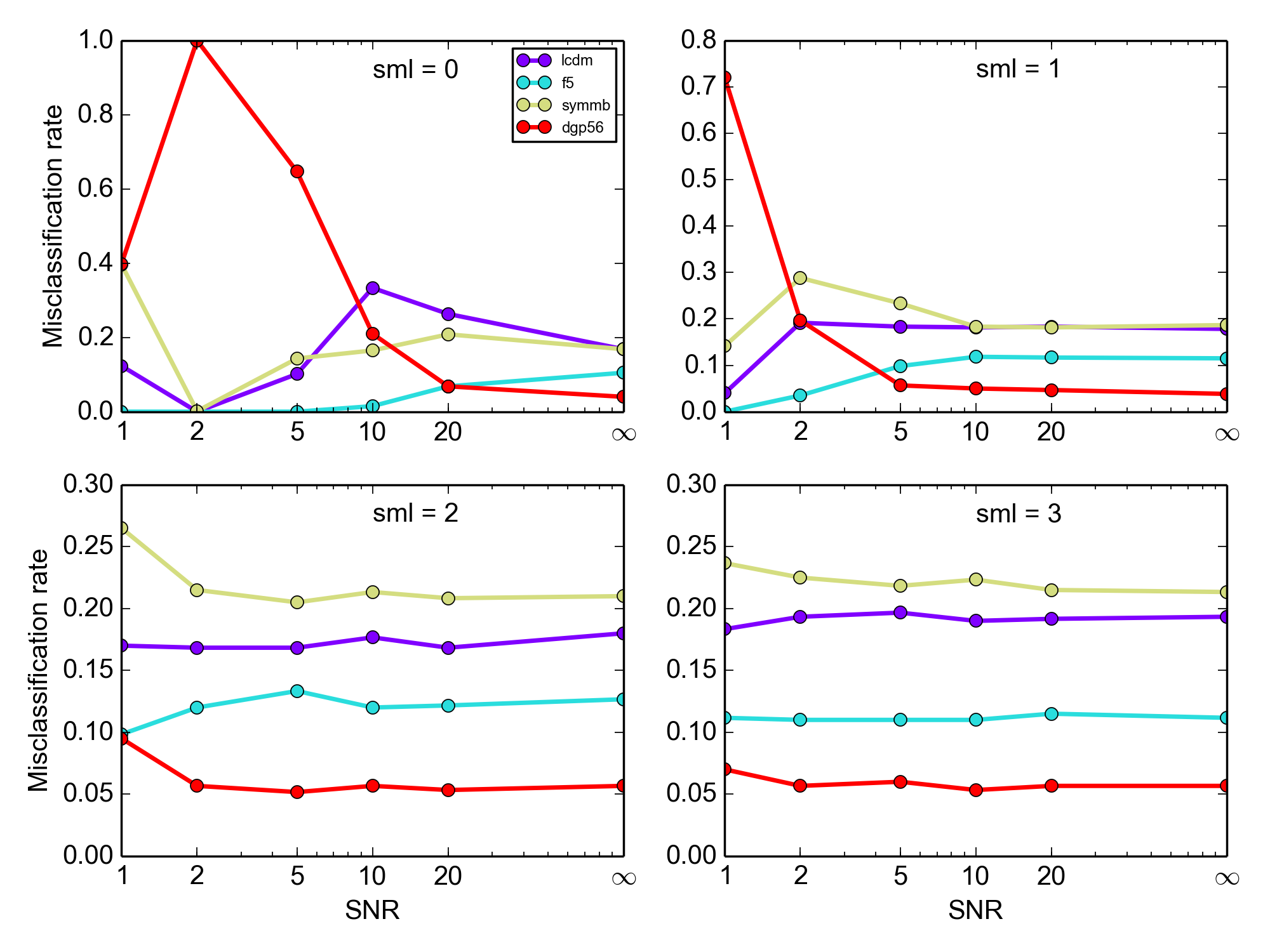}
\caption{The mis-identification rate for all models as a function of the same
parameters as in figure \ref{FIG::DIS::MODEL_SUCCESS}.}
\label{FIG::DIS::MODEL_MIS}
\end{figure}

In order to do so, we show in figure \ref{FIG::DIS::MODEL_SUCCESS} the
success rate per model as a function of signal to noise ratio (SNR) and with 
different
levels of smoothing applied. The bottom two panels of the figure, with strong
smoothing, show the typical success rate hierarchy of models, which is largely
constant over the full SNR range. With no or weak smoothing
applied, however, a pattern emerges. The success rate for \textit{dgp56} peaks
sharply with an ideal success rate at $\textrm{SNR}=2$, which slightly
drops for the lowest SNR images. The success rate of all other
models vanishes at this \textit{dgp56} SNR sweet spot. This trend
is clarified by figure \ref{FIG::DIS::MODEL_MIS}, which, as a function of the
same parameters shows the relative mis-classification rate per model. The
combination of the two figures tells us that at $\textrm{SNR}=2$
and without smoothing, every image gets classified as \textit{dgp56}. This
trends shifts to $\textrm{SNR}=1$ if we apply a one pixel
smoothing. This suggests that image features of the \textit{dgp56} training set
can be mimicked by noise in the test images. This is good and bad. It is good
because it already hints towards that \textit{dgp56} images seem to contain a
lot of small-scale structure, but it is also very alarming since it clearly
shows that our analysis can be largely biased towards specific models in the
presence of noise. Luckily, the described effects can be easily alleviated by
the application of smoothing. However, our noise model was particularly simple
and the addition of more realistic noise might give way to more subtle effects.

\subsection{Dependence on $\sigma_{8}$}
\label{SEC::DIS::SIGMA8}

All of our simulations have exactly the same initial 
conditions. One consequence of this is that the amplitude of matter 
fluctuations today, $\sigma_8$, will be different: for \textit{lcdm} we have 
$\sigma_8 = 0.8$, for \textit{dgp12}, \textit{f5} and \textit{symmB} we have 
$\sigma_8 = 0.85$ while for \textit{dgp56}, \textit{f6} and \textit{symmA} we 
have $\sigma_8 = 0.82$. \textit{dgp05} produces $\sigma_{8} = 0.89$.

The difference in structure formation between modified gravity 
models and $\Lambda$CDM will be degenerate with a change of this amplitude, 
i.e. we can mimic some of the effects of a modified gravity model at a given 
redshift by simply running a $\Lambda$CDM simulation with a higher $\sigma_8$. 
This is especially true for the DGP models for which the modifications of 
gravity, in the absence of screening, is equivalent to a time-dependent 
Newtonian constant $G_{\rm eff}(a) = G(1 + C(a))$. The presence of a screening 
mechanism, which leads to general relativity being recovered inside the most 
massive haloes, will act to reduce this degeneracy.

\begin{figure}
\includegraphics[width=.5\textwidth]{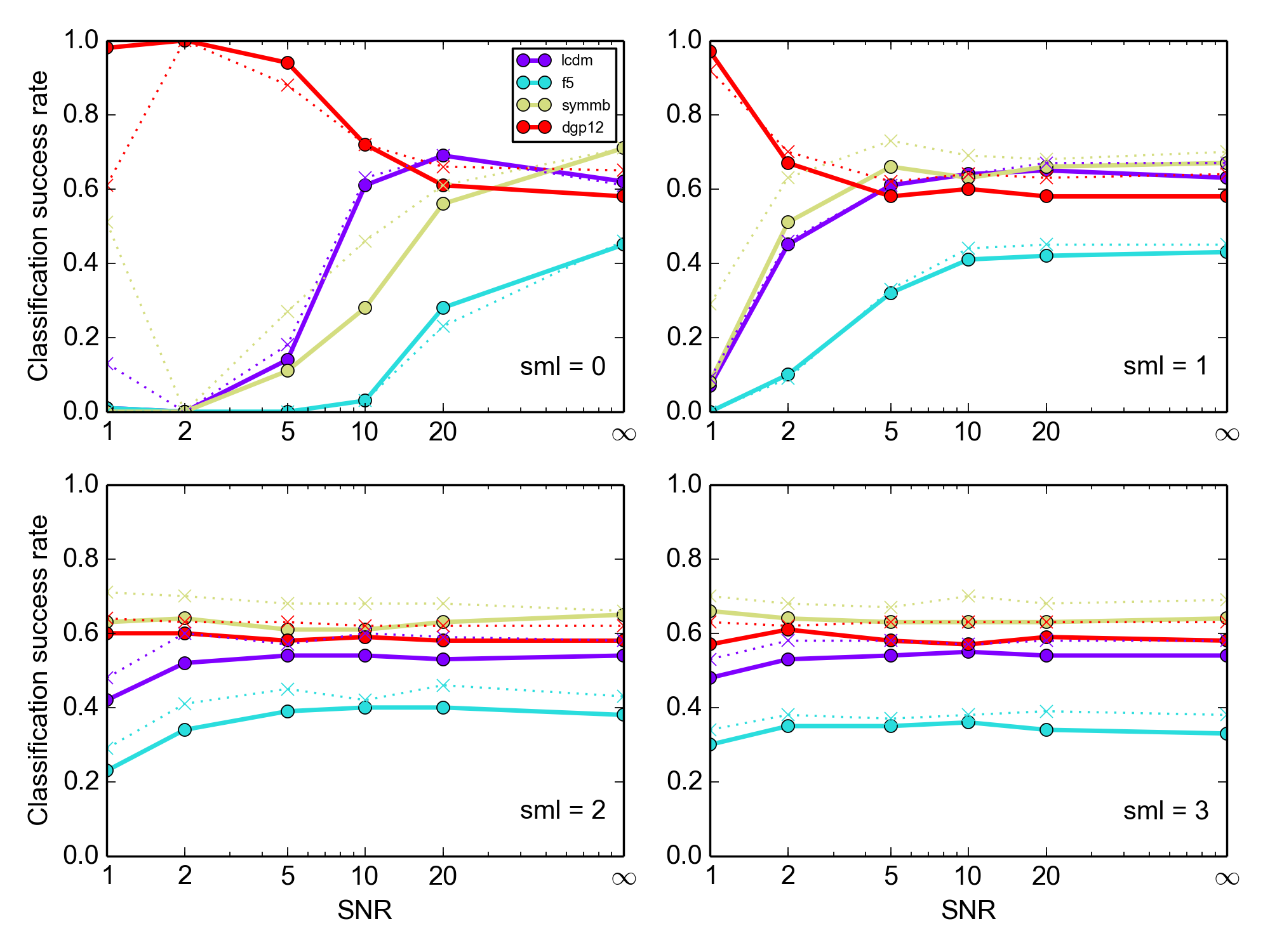}
\caption{This figure is the same as figure
\ref{FIG::DIS::MODEL_SUCCESS}, but with the \textit{dgp56} model switched in
place with a \textit{dgp12} model. For comparison, the former results of figure
\ref{FIG::DIS::MODEL_SUCCESS} are shown by the thinner, dotted lines.}
\label{FIG::DIS::DGP12}
\end{figure}

We investigated the response of the classification results to different values
of $\sigma_{8}$ by running a new \textit{lcdm} simulation with $\sigma_{8}
= 0.85$ instead of $\sigma_{8} = 0.8$. We then directly compare the two
\textit{lcdm} models and show the classification matrices in table
\ref{TAB::DIS::SIGMA8}. The classification is indeed quite
sensitive to this change in parameters. The result is confirmed in table
\ref{TAB::DIS::SIGMA8_SS}, for a lower $\textrm{SNR}=5$
and with a two pixel smoothing applied. On the one hand, this is very
encouraging since it shows that the method is sensitive to changes in
cosmological parameters for the same model of gravity. Hence, the
computer vision approach can be used to constrain cosmological parameters
from observations when using numerical simulations on a parameter grid
\citep[e.g.][for a recent suitable simulation suite]{Heitmann2016} as
training sets. But on the other hand, it also shows us that some some effects 
we were seeing
earlier indeed stem from a different $\sigma_{8}$ seen in the different models. 
We therefore run another set of classifications, where we exchange the
\textit{dgp56} model with the \textit{dgp12} model, which produces a different
$\sigma_{8}$. We show the results in table \ref{TAB::DIS::DGP12} and figure
\ref{FIG::DIS::DGP12}. There we see that the overall classification success rate
is slightly worse, due to the fact that \textit{dgp12} is slightly more
degenerate with other models. However, all major trends that we have seen
earlier are effectively unchanged, which suggests that the change in
$\sigma_{8}$ between the simulations is indeed not the main driver for the 
ability to distinguish the models. A similar classification as the one 
underlying table \ref{TAB::DIS::DGP12} is shown in appendix \ref{SEC::APP}, 
where the baseline \textit{lcdm} model is switched in place with the one the 
produces the higher $\sigma_{8}$.

\begin{table}
\caption{Direct model on model comparison for two \textit{lcdm}
models, with different $\sigma_{8}$ parameters.}
\label{TAB::DIS::SIGMA8}
\begin{tabular}{l|ccc|}
\hline
&$\sigma_{8}=0.8$&$\sigma_{8}=0.85$&success rate\\
\hline
       &192&8&\\
$\sigma_{8}=0.8$ &0.7829&0.2171&0.96\\
	 &1.0&0.28&\\
\hline
 	&56&144&\\
$\sigma_{8}=0.85$  &0.2515&0.7485&0.72\\
	&0.34&1.00&\\
\hline
\hline
all &&&0.84
\end{tabular}
\end{table}

\begin{table}
\caption{The same comparison as table
\ref{TAB::DIS::SIGMA8} but for $\textrm{SNR}=5$ and a smoothing scale of 2 
pixels.}
\label{TAB::DIS::SIGMA8_SS}
\begin{tabular}{l|ccc|}
\hline
&$\sigma_{8}=0.8$&$\sigma_{8}=0.85$&success rate\\
\hline
       &187&13&\\
$\sigma_{8}=0.8$ &0.7722&0.2278&0.94\\
	 &1.0&0.29&\\
\hline
 	&60&140&\\
$\sigma_{8}=0.85$  &0.3016&0.6984&0.70\\
	&0.43&1.00&\\
\hline
\hline
all &&&0.82
\end{tabular}
\end{table}

\begin{table}
\caption{Classification matrices with the \textit{dgp12} model.}
\label{TAB::DIS::DGP12}
\begin{tabular}{l|ccccc|}
\hline
&lcdm&f5&symmB&dgp12&success rate\\
\hline
       &123&27&42&8\\
lcdm &0.3537&0.2556&0.2530&0.1377&0.62\\
	 &1.00&0.72&0.72&0.39&\\
\hline
 	&36&90&68&6&\\
f5  &0.2555&0.3424&0.2900&0.1121&0.45\\
	&0.75&1.00&0.85&0.33&\\
\hline
		&29&24&141&6&\\
symmB	&0.2514&0.2767&0.3596&0.1124&0.71\\
		&0.70&0.77&1.00&0.31&\\
\hline
 		&46&13&25&116\\
dgp12   &0.1799&0.1498&0.1544&0.5160&0.58\\
		&0.35&0.29&0.30&1.00&\\
\hline
\hline
all &&&&&0.59
\end{tabular}
\end{table}

\subsection{Image features}
\begin{figure*}
\includegraphics[width=\textwidth]{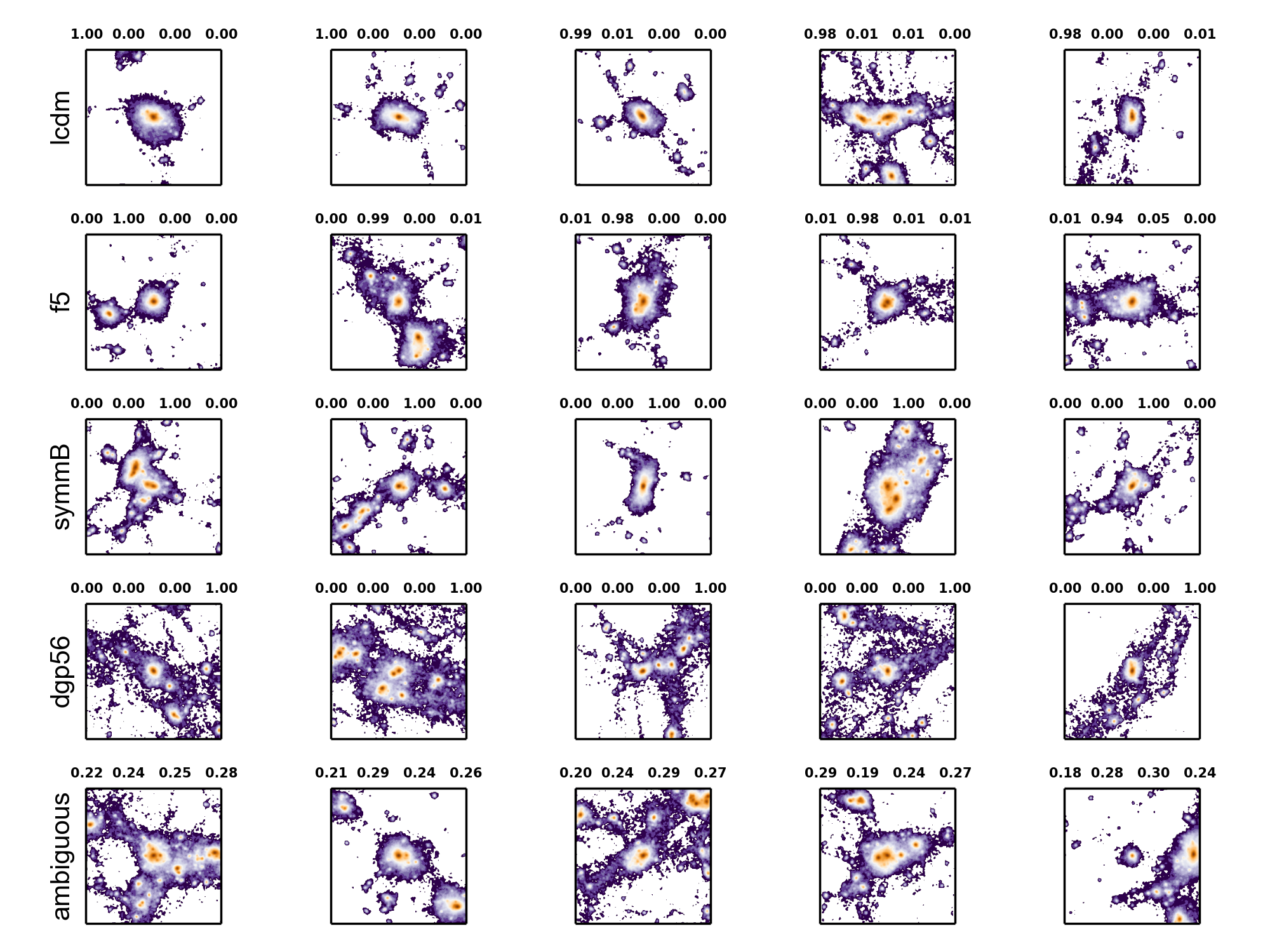}
\caption{Test images with peculiar similarity values. The first four rows show
the five images per model class with the highest similarity to their true model
class. The numbers above each thumbnail image show similarity value to the
\textit{lcdm}, \textit{f5}, \textit{symmB} and \textit{dgp56} class of 
the training set,
respectively. The bottom row shows the five most ambiguous test images, meaning
that they are classified as similarly close to all training set classes.}
\label{FIG::DIS::THUMBNAILS}
\end{figure*}

To look more into the specific images features of each model, we take a closer
look at the similarity parameter defined by equation \ref{EQ::METHOD::SIM} for
individual test images. In the first four rows of figure
\ref{FIG::DIS::THUMBNAILS} we show those five test images per class that
scored the highest similarity values to their true model, respectively. The
bottom row shows the most ambiguous images, where the distribution of 
similarities to all classes is most uniform\footnote{The five 
images were chosen such to have the minimum variance in their 
four similarity values.}. The images in
the four top rows can hence we interpreted as prototype images, best
representing the features of their respective training set class. For the
\textit{lcdm} model we see mostly isolated haloes with a quite elliptical core,
surrounded by a number of small satellites. The fourth \textit{lcdm}
image in figure \ref{FIG::DIS::THUMBNAILS} shows a merger, where each merging
component in the field is again a elliptical core halo with a number of
satellites. At least by visual inspection, the typical \textit{f5} halo has a
rounder core and is surrounded by more massive substructure. The same is true
for \textit{symmB} but the overall structure of the images looks more
complicated, with either very elliptical or extremely complex cores which can
show several peaks. As expected from the earlier discussion, the 
\textit{dgp56} images look most distinct with very pronounced filamentary 
structure around the halo and many coarse, low-intensity features along these 
filaments. We assume that it is this filamentary structure which can be easily 
mimicked by noise and bias the classification towards this model. The most 
ambiguous haloes are usually some special situations where we have several very 
massive and distinct structures on a single field or particularly complicated 
merger situations.

Obviously, the descriptions above are very subjective and may miss subtle
features. In fact our aim to use computer vision based classifiers is motivated
in avoiding such descriptions but to quantitatively characterise 
images. The interpretation of 2D halo morphology  for 
the modified gravity models of table  using computer vision is  listed in
table \ref{TAB::DIS::FEATURES}, which shows the 25 image features with the 
highest fisher discriminants for the classification run summarised in table 
\ref{TAB::DIS::OPTIMAL}. This particular list is derived from training sets 
smoothed on a one pixel scale, which produced the best results, but we have 
verified that there are only marginal changes in feature fisher scores between 
different smoothing 
scales. For completeness we list the remaining 48 features that were used in 
this analysis in appendix \ref{APP::LIST}. Some of the top-ranked features are
easy to interpret in terms of halo morphology. For example, the first Haralick
texture is the angular second moment, which is a measure of the complexity of
an image. The fact that this feature is most descriptive on the Edge transform
of the image is not surprising since this transformation just filters out the
actual structure in the image and hence reduces the noise. However, most of the
other main features are more difficult to interpret directly, such as the
Zernike-coefficients of the Fourier transform, the Zernike-coefficients of the
Fourier transform of the Wavelet transform and Tamura textures of the Chebyshev
transform. Another very discriminating feature seems to be the Gini coefficient
for different transformations, another measure that describes the homogeneity of
an image. To get a bigger picture we list in table \ref{TAB::DIS::FAMILIES} the
15 most significant feature classes ordered by the mean
fisher score within the class. Also minimum, maximum and standard deviation
of the fisher scores within the family are given in the table. Again, we see
that the Gini coefficient for different transformations, together with Haralick
and Tamura textures and Zernike coefficients are most relevant. One should not
over-interpret the actual ranking within that list though since the number of
features in each family is varying strongly. For example, the single features
with the highest Fisher score belong to the families of Haralick
textures and Zernike coefficients, but they are hidden within the mean of the 
feature family in this table. While it would be excessive to list the fisher 
scores for all 2919 features\footnote{Full classification data available upon 
e-mail request from the authors.}, but we list in appendix \ref{APP::LIST} the 
fisher score statistics of the remaining 109 feature families.

\begin{table}
\caption{The most discriminating image features according to their fisher 
discriminant.}
\begin{tabular}{ccc}
\hline
Rank & Name & Weight \\
\hline
\hline
1 & Haralick Textures (Edge ()) [0] & 0.4908 \\
2 & Zernike Coefficients (Fourier ()) [19] & 0.4173 \\
3 & Zernike Coefficients (Fourier ()) [26] & 0.4091 \\
4 & Zernike Coefficients (Fourier ()) [41] & 0.402 \\
5 & Zernike Coefficients (Fourier ()) [54] & 0.3961 \\
6 & Zernike Coefficients (Fourier ()) [28] & 0.3958 \\
7 & Tamura Textures (Chebyshev ()) [2] & 0.3917 \\
8 & Haralick Textures (Edge ()) [22] & 0.3881 \\
9 & Haralick Textures (Edge ()) [18] & 0.3838 \\
10 & Zernike Coefficients (Fourier ()) [5] & 0.3779 \\
11 & Zernike Coefficients (Fourier (Wavelet ())) [19] & 0.3723 \\
12 & Zernike Coefficients (Fourier ()) [71] & 0.372 \\
13 & Zernike Coefficients (Fourier (Wavelet ())) [28] & 0.3718 \\
14 & Gini Coefficient (Fourier (Wavelet ())) [0] & 0.3661 \\
15 & Zernike Coefficients (Fourier ()) [69] & 0.3655 \\
16 & Zernike Coefficients (Fourier (Wavelet ())) [69] & 0.3648 \\
17 & Haralick Textures (Edge ()) [10] & 0.3634 \\
18 & Zernike Coefficients (Fourier (Wavelet ())) [54] & 0.3618 \\
19 & Zernike Coefficients (Fourier (Wavelet ())) [5] & 0.3612 \\
20 & Zernike Coefficients (Fourier (Wavelet ())) [41] & 0.357 \\
21 & Zernike Coefficients (Fourier (Wavelet ())) [39] & 0.3553 \\
22 & Gini Coefficient (Fourier ()) [0] & 0.3538 \\
23 & Haralick Textures (Edge ()) [15] & 0.3508 \\
24 & Zernike Coefficients (Fourier ()) [67] & 0.3482 \\
25 & Haralick Textures () [4] & 0.3462 \\
\end{tabular}
\label{TAB::DIS::FEATURES}
\end{table}

\begin{table*}
\caption{Fisher discriminant statistics of feature classes ordered by the mean.}
\begin{tabular}{ccccccc}
\hline
Rank & Name & Min & Max & Mean & Std. dev. & \# \\
\hline
\hline
1 & Gini Coefficient (Fourier (Wavelet ())) & 0.3661 & 0.3661 & 0.3661 & -- &1\\
2 & Gini Coefficient (Fourier ()) & 0.3538 & 0.3538 & 0.3538 & -- &1\\
3 & Gini Coefficient (Wavelet (Fourier ())) & 0.259 & 0.259 & 0.259 & -- &1 \\
4 & Gini Coefficient (Chebyshev (Wavelet ())) & 0.2485 & 0.2485 & 0.2485 & --&1
\\
5 & Gini Coefficient (Edge ()) & 0.2361 & 0.2361 & 0.2361 & --&1 \\
6 & Gini Coefficient (Wavelet (Edge ())) & 0.2178 & 0.2178 & 0.2178 & --&1 \\
7 & Tamura Textures (Chebyshev ()) & 0.01879 & 0.3917 & 0.2066 & 0.151&6 \\
8 & Haralick Textures () & 0.001923 & 0.3462 & 0.153 & 0.1126&28 \\
9 & Haralick Textures (Edge ()) & 0.002559 & 0.4908 & 0.1489 & 0.154&28 \\
10 & Zernike Coefficients (Fourier ()) & 0.001196 & 0.4173 & 0.1403 & 0.1268&72
\\
11 & Zernike Coefficients (Fourier (Wavelet ())) & 0.0004068 & 0.3723 & 0.138 &
0.1139&72 \\
12 & Gini Coefficient () & 0.1269 & 0.1269 & 0.1269 & --&1 \\
13 & Tamura Textures (Chebyshev (Wavelet ())) & 0.0 & 0.2066 & 0.1167 &
0.09447&6 \\
14 & Gini Coefficient (Chebyshev ()) & 0.1046 & 0.1046 & 0.1046 & --&1 \\
15 & Tamura Textures (Fourier ()) & 0.0 & 0.2337 & 0.083 & 0.1123&6 \\
\end{tabular}
\label{TAB::DIS::FAMILIES}
\end{table*}


 \section{Summary and outlook}
 \label{SEC::CONCL}
This work introduces computer vision characterisation of 2D images of dark
matter surface mass density distributions to classify different models of
structure formation. We used a set of numerical simulations, all run with
identical initial conditions, but with different underlying models of gravity. 
Our
simulations included a standard $\Lambda$CDM model, two $f(R)$ models, two
Symmetron models and three DGP models. We extracted the 200 most massive
haloes in each of the simulation boxes and produced 4000 images of the
surface mass density distribution along 20 different lines-of-sight. On these
images we ran a computer vision characterisation algorithm, which extracts up
to 2919 image features and then used a simple WND and WNN based classification
algorithm to classify a set of 200 test images for each class of models. Both,
the computer vision characterisation algorithm and the subsequent classifier
are implemented in the publicly available software package \texttt{WND-CHARM}.
To the test images we applied several levels of uncorrelated noise and probed
the effects of smoothing on both, the training and test images.

There are degeneracies between certain models when using image morphology for
classification. Especially the DGP models are strongly degenerate between
themselves, while Symmetron and $f(R)$ models show only mild deviations from
$\Lambda$CDM. To quantify these degeneracies we ran model-on-model
classifications looping through every possible model pair and report the
results in section \ref{SEC::RESULTS}. After this test, we reduced the number
of models to four, with only a single model per gravity family. Using the
classifier now on these four models shows a typical classification success rate
of about 60\%. While this is a good success rate, it is not 
robust enough to accurately classify single haloes. We want to highlight,
however, that each class of test images was robustly classified correctly as an 
ensemble,
as shown in table \ref{TAB::RES::MAIN}. This result even holds when using the 
full set of eight models as shown in 
table
\ref{TAB::RES::SUCCESS_RATE::APP}. As expected, the method is quite sensitive
to the presence of noise, which in our case was just white pixel noise, applied
to the test images. The classification success rates drop to the level
of 30\% as shown in figure \ref{FIG::RES::SML_PARAM_SNR}, which is only
slightly better than a characterisation by random pick. But figure 
\ref{FIG::RES::SML_PARAM_SNR} also shows that this
sensitivity to noise can easily be remedied by the application of mild
smoothing to the training and test images, while only marginally affecting the
overall classification success rate compared to the ideal case where no noise
is present and no smoothing is applied.

Several strategies can be applied to increase the classification success rate,
e.g. by smoothing the training sets and restricting the classification
to only a subset of the top fisher score weighted characterisation features.
Another possibility is to use additional information on the halo images, for
example the a rough estimate of the  mass of the object. With such 
techniques,
overall classification success rates can exceed 70\%. While the classification
accuracy, as expected, decreases for larger redshifts (table
\ref{TAB::RES::REDSHIFT_BINS}), our methodology can be used to classify haloes
into redshift classes based on their morphology only (table \ref{TAB::RES::Z}). 
Since all our simulations where run with the same initial conditions, different 
gravity models may produce different amplitudes of the density fluctuations 
today, usually expressed in terms of the parameter $\sigma_{8}$. We 
have extensively tested that our classification is not solely driven by 
potential 
variations of $\sigma_{8}$ in the probed models. The approach is however 
sensitive to these variations and it could be used to constrain it from 
observations. These results are summarised in section \ref{SEC::DIS::SIGMA8} 
and appendix \ref{SEC::APP}. 

The presence of noise can trigger strong biases and prefer the classification
of certain models over others as we show in section \ref{SEC::DIS}. Figure
\ref{FIG::DIS::THUMBNAILS} shows that the DGP models tend to produce low pixel
intensity filamentary structure around the more massive haloes. This
morphological feature can easily be mimicked by the addition of noise and is
then incorrectly picked up as a feature by the classifier. Although this is 
easily mended by the introduction of suitable
smoothing, it clearly shows that some models are more susceptible to noise than
others. These findings also highlight that the morphology of the different
models indeed differ significantly, firstly shown by the ability of our
classifier to distinguish them and visually highlighted by figure
\ref{FIG::DIS::THUMBNAILS}, which shows the most typical prototypes of a
certain class, as well as the most ambiguous haloes in the test sample.

As the main finding of this work, we present the most discriminating image
features to classify the halo images. Their individual and average feature
class fisher discriminant scores are reported in tables \ref{TAB::DIS::FEATURES}
and \ref{TAB::DIS::FAMILIES}, as well as in appendix \ref{APP::LIST}. Texture
features such as the Tamura and Haralick textures (see section
\ref{SEC::METHOD::CHAR} for a description) are among the most informative
features for classification. While for the Haralick textures, the actual image
data or the similar Edge transform is the most descriptive starting point, the
Tamura textures are more descriptive in the, to the human perception, more
abstract Fourier -or Chebyshev domain. The same is true for the other most
descriptive features, being the coefficients of the Zernike decomposition of
the image either in Fourier space, or based on the Fourier transform of the
Wavelet transform. Finally, the Gini coefficient seems to be a valuable feature
in terms of classification when applied to any kind of pixel intensity
transformation. In conclusion, this study shows that 2D images of the dark
matter distribution can indeed be used to identify the signatures of different
models of structure formation, in our case the signatures of different models
of gravity. This is most promising since techniques such as gravitational
lensing mass reconstructions can deliver such images from observations.

This work is only a first attempt of a computer vision based classification of
different models of structure formation. We set it up as an initial proof of
concept of the feasibility of this approach. Results seem promising, but we
have to point out that our simplistic noise model does not capture all of the
complications that a real study of lensing surface mass density maps would
entail. The effects of the inclusion of correlated noise deriving from
large-scale structure along the line-of-sight \citep[among
others][]{Becker2011,Hoekstra2011} needs to be assessed. Furthermore, the
systematic effects in the reconstruction of a surface mass density map from
lensing data must be mimicked. Such frameworks exist
\citep{Meneghetti2010a,Meneghetti2016} and shall be a next step in this study.
It also needs to be verified if such surface mass density maps are indeed
needed at all. Computer vision classification could potentially
work directly in shear space, a direct lensing observable from galaxy surveys
\citep{Bartelmann2001} and a quantity that can be derived, although under more
difficulty,  from numerical simulations \citep[e.g.][]{Barreira2016}. Also 
our 
current classifier based on either WND or WNN distance measures is potentially 
not the ideal solution. More sophisticated techniques based on machine learning 
could
be applied to the most descriptive feature families, potentially after a
dimensionality reduction via a principal component analysis. The overall
approach can potentially  be extended to much wider fields, probing the
morphology of the cosmic web itself and not only of its nodes. With on-going
and upcoming wide-field or all-sky galaxy surveys, the necessary input data
will be available, but further tests with simulations must verify the 
robustness of the approach.
Finally, in this work we focused only on models of modified gravity to be
classified. The halo characterisation framework, however, is fully general and 
can readily be extended to the classification of different models of dark 
matter, dark energy interactions, baryonic physics or different 
global cosmological setups, as we already hinted towards in section 
\ref{SEC::DIS::SIGMA8}.

\section*{Acknowledgements}
Large parts of this work have been carried out during an Astrophysics summer 
research programme within the Department of Physics at the University of 
Oxford. We would like to thank the organisers for this excellent research 
opportunity for young students, which enabled QL's substantial contribution to 
this work. We also want to thank the authors of the excellent 
\texttt{WND-CHARM} algorithm for the development and open access to this 
software 
package. JM has received funding from the Marie Curie 
Actions  of the European Union's Horizon 2020 Programme under REA grant 
agreement number 627288 (WEBMAP). HAW 
is supported by the European Research Council through 646702 (CosTesGrav).

\input{mg_computer_vision.bbl}

\appendix
\section{Full set of models}
\label{SEC::APP}
This appendix shows some classification results if all eight modified 
gravity models are used in
the analysis. Table \ref{TAB::RES::SUCCESS_RATE::APP} shows the same
classification success matrices as the four model equivalent in table
\ref{TAB::RES::MAIN}.
Although the overall classification success rate and the results for each model
are now significantly worse, since no degeneracies are removed, it should
be pointed out that all model success rates are well above 12.5\%, the
expectation for a random pick classification. Furthermore, for all models
besides one exception, the correct model is clearly preferred over the others.
The exception is \textit{symmA}, which by raw numbers has more \textit{f6}
images associated to it then from its correct underlying class . However, also 
this trend is remedied once one considers
the average similarity per class.

Physically interesting results can be seen in table
\ref{TAB::RES::REDSHIFT_BINS::APP}, which shows the model classification
success rates as a function of redshift.  Two results catch the eye. Firstly,
\textit{symmA} is not classified at all at $z=1$, which relates to the fact
that this specific gravity model has no signature at this high redshift. Other
degeneracies fully dominate in this case.
Secondly, \textit{dgp05} has a particularly high classification success rate at
$z=0.5$. This effect is not quite clear to us and we have to postpone its 
explanation to future studies. Finally, as observed 
earlier,
classification success rate usually go down as a function of redshift, at it
was already discussed in section \ref{SEC::DIS}.

Some tendencies can be understood from looking at other 
commonly studied observables. For example the similarities we see between 
\textit{f6} and \textit{symmA} is not that surprising given that these two 
models are also close when considering classical clustering observables. For 
example the matter power spectrum and halo mass function at $z=0$ is very close 
in these two models.

To show a comparison of models which have all a similar value 
of $\sigma_{8}$ (compare section \ref{SEC::DIS::SIGMA8}), we show in table 
\ref{TAB::DIS::DGP12_LCDM_R} the classification matrices for training and test 
sets that contain \textit{f5}, \textit{symmB}, \textit{dgp12} and a 
\textit{lcdm} model that produces $\sigma_{8}=0.85$. The results agree well 
qualitatively with what has been seen earlier, but shift some of the 
degeneracies.

\begin{table*}
\caption{Classification success and similarity matrix}
\label{TAB::RES::SUCCESS_RATE::APP}
\begin{tabular}{l|ccccccccc|}
\hline
&lcdm&f5&f6&symmA&symmB&dgp05&dgp12&dgp56&success rate\\
\hline
       &73&20&33&30&33&4&5&2&\\
lcdm &0.1876&0.1364&0.1578&0.1638&0.1378&0.0771&0.0708&0.0686&0.37\\
	 &1.00&0.73&0.84&0.87&0.73&0.41&0.38&0.37\\
\hline
 	&18&83&17&14&58&5&3&2&\\
f5  &0.1452&0.1945&0.1522&0.1512&0.1639&0.0664&0.0639&0.0626&0.42\\
	&0.75&1.00&0.78&0.78&0.84&0.34&0.33&0.32\\
\hline
 	&32&19&52&39&42&5&8&3&\\
f6	&0.1579&0.1464&0.1650&0.1769&0.1528&0.0704&0.0664&0.0644&0.26\\
	&0.89&0.82&1.00&0.95&0.88&0.45&0.41&0.40\\
\hline
	 	&33&23&48&45&40&3&4&4&\\
symmA	&0.1579&0.1464&0.1650&0.1769&0.1528&0.0704&0.0664&0.0644&0.23\\
		&0.89&0.83&0.93&1.00&0.86&0.40&0.38&0.36&\\
\hline
		&16&21&24&15&117&3&4&0&\\
symmB	&0.1457&0.1536&0.1532&0.1580&0.2000&0.0648&0.0641&0.0607&0.59\\
		&0.73&0.77&0.77&0.79&1.00&0.32&0.32&0.30& \\
\hline
		&24&9&18&10&14&66&36&23&\\
dgp05	&0.0876&0.0769&0.0875&0.0848&0.0794&0.2164&0.1908&0.1766&0.33\\
		&0.40&0.36&0.40&0.39&0.37&1.00&0.88&0.82&\\
\hline
		 &24&9&16&15&17&31&56&32&\\
dgp12	 &0.0891&0.0772&0.0899&0.0880&0.0815&0.1839&0.2034&0.1871&0.28\\
		 &0.44&0.38&0.44&0.43&0.40&0.90&1.00&0.92&\\
\hline
 		&22&5&23&8&12&28&34&68&\\
dgp56   &0.0827&0.0711&0.0845&0.0785&0.0720&0.1813&0.1953&0.2347&0.34\\
		&0.35&0.30&0.36&0.33&0.31&0.77&0.83&1.00&\\
		
\hline
\hline
all&&&&&&&&&0.35
\end{tabular}
\end{table*}

\begin{table}
\caption{Classification success rate at different redshifts}
\label{TAB::RES::REDSHIFT_BINS::APP}
\begin{tabular}{lccc}
\hline
z&0.0&0.5&1.0\\
\hline
\hline
lcdm&0.37&0.27&0.29\\
\hline
f5&0.42&0.47&0.39\\
\hline
f6&0.26&0.24&0.26\\
\hline
symmA&0.23&0.25&\textbf{0.0}\\
\hline
symmB&0.59&0.40&0.31\\
\hline
dgp05&0.33&\textbf{0.82}&0.24\\
\hline
dgp12&0.28&0.29&0.20\\
\hline
dgp56&0.34&0.47&0.31\\
\hline
\hline
total&0.35&0.40&0.25\\
\end{tabular}
\end{table}

\begin{table}
\caption{Classification matrices with the \textit{dgp12} model and with a 
$\textit{lcdm}_{0.85}$ model with $\sigma_{8}=0.85$.}
\label{TAB::DIS::DGP12_LCDM_R}
\begin{tabular}{l|ccccc|}
\hline
&$\textit{lcdm}_{0.85}$&f5&symmB&dgp12&success rate\\
\hline
       &111&15&36&38\\
$\textit{lcdm}_{0.85}$&0.4694&0.1115&0.1252&0.2939&0.56\\
	 &1.00&0.24&0.27&0.63&\\
\hline
 	&2&112&80&6&\\
f5  &0.1003&0.4063&0.3532&0.1402&0.56\\
	&0.25&1.00&0.87&0.35&\\
\hline
		&2&45&147&6&\\
symmB	&0.1032&0.3353&0.4209&0.1406&0.74\\
		&0.25&0.80&1.00&0.33&\\
\hline
 		&32&30&47&91\\
dgp12   &0.2807&0.1763&0.1846&0.3583&0.46\\
		&0.78&0.49&0.52&1.00&\\
\hline
\hline
all &&&&&0.58
\end{tabular}
\end{table}

\section{Full fisher score statistics for feature families}
\label{APP::LIST}
It would be excessive to list the fisher discriminant score for each of 2919
features in this image characterisation exercise. But to show at least the
complete feature vector that was used in the analysis leading to section
\ref{SEC::DIS::OPTIMAL} we complete table \ref{TAB::DIS::FEATURES} with table
\ref{TAB::DIS::FEATURES_APP} showing the remaining features with their Fisher
scores in the total feature vector with 73 elements.

Table \ref{TAB::DIS::FAMILIES_APP} completes table
\ref{TAB::DIS::FAMILIES} and now lists all 124 image feature classes that
contain the 2919 image features, together with basis statistics on the Fisher
scores within the class.

\begin{table}
\caption{Table \ref{TAB::DIS::FEATURES} continued}
\begin{tabular}{ccc}
\hline
Rank & Name & Weight \\
\hline
\hline
26 & Tamura Textures (Chebyshev ()) [5] & 0.3426 \\
27 & Zernike Coefficients (Fourier (Wavelet ())) [71] & 0.3379 \\
28 & Haralick Textures () [14] & 0.3273 \\
29 & Haralick Textures (Edge ()) [14] & 0.3211 \\
30 & Haralick Textures (Wavelet ()) [4] & 0.3202 \\
31 & Zernike Coefficients (Fourier (Wavelet ())) [65] & 0.3156 \\
32 & Haralick Textures (Edge ()) [11] & 0.3154 \\
33 & Multiscale Histograms (Chebyshev (Fourier ())) [22] & 0.3119 \\
34 & Zernike Coefficients (Fourier (Wavelet ())) [10] & 0.3094 \\
35 & Haralick Textures () [10] & 0.306 \\
36 & Haralick Textures () [8] & 0.3028 \\
37 & Haralick Textures () [18] & 0.2981 \\
38 & Zernike Coefficients (Fourier ()) [62] & 0.292 \\
39 & Multiscale Histograms (Chebyshev (Fourier ())) [21] & 0.2874 \\
40 & Zernike Coefficients (Fourier ()) [48] & 0.287 \\
41 & Zernike Coefficients (Fourier ()) [39] & 0.2864 \\
42 & Haralick Textures (Wavelet (Edge ())) [4] & 0.2854 \\
43 & Haralick Textures (Edge ()) [9] & 0.2821 \\
44 & Zernike Coefficients (Fourier ()) [24] & 0.2801 \\
45 & Haralick Textures (Edge ()) [8] & 0.2789 \\
46 & Zernike Coefficients (Fourier ()) [34] & 0.2762 \\
47 & Haralick Textures () [22] & 0.2688 \\
48 & Haralick Textures (Fourier (Chebyshev ())) [18] & 0.2683 \\
49 & Zernike Coefficients (Fourier ()) [57] & 0.265 \\
50 & Zernike Coefficients (Fourier (Wavelet ())) [37] & 0.2646 \\
51 & Zernike Coefficients (Fourier ()) [8] & 0.2602 \\
52 & Gini Coefficient (Wavelet (Fourier ())) [0] & 0.259 \\
53 & Haralick Textures () [20] & 0.258 \\
54 & Haralick Textures (Fourier (Chebyshev ())) [12] & 0.2578 \\
55 & Zernike Coefficients (Fourier (Wavelet ())) [34] & 0.2557 \\
56 & Zernike Coefficients (Fourier (Wavelet ())) [24] & 0.2533 \\
57 & Zernike Coefficients (Fourier ()) [45] & 0.2526 \\
58 & Haralick Textures (Wavelet (Edge ())) [12] & 0.2517 \\
59 & Haralick Textures () [19] & 0.2517 \\
60 & Zernike Coefficients (Fourier (Wavelet ())) [62] & 0.2514 \\
61 & Haralick Textures () [1] & 0.2502 \\
62 & Haralick Textures (Chebyshev (Wavelet ())) [14] & 0.2495 \\
63 & Gini Coefficient (Chebyshev (Wavelet ())) [0] & 0.2485 \\
64 & Zernike Coefficients (Fourier (Wavelet ())) [48] & 0.2463 \\
65 & Zernike Coefficients (Fourier (Wavelet ())) [8] & 0.2461 \\
66 & Haralick Textures (Chebyshev ()) [8] & 0.2406 \\
67 & Haralick Textures (Chebyshev (Wavelet ())) [10] & 0.2401 \\
68 & Gini Coefficient (Edge ()) [0] & 0.2361 \\
69 & Tamura Textures (Fourier ()) [5] & 0.2337 \\
70 & Zernike Coefficients (Fourier (Wavelet ())) [46] & 0.2327 \\
71 & Haralick Textures (Chebyshev ()) [26] & 0.2326 \\
72 & Haralick Textures (Chebyshev ()) [2] & 0.2308 \\
73 & Zernike Coefficients (Fourier (Wavelet ())) [31] & 0.2299 \\
\end{tabular}
\label{TAB::DIS::FEATURES_APP}
\end{table}

\begin{table*}
\caption{Table \ref{TAB::DIS::FAMILIES} continued}
\label{TAB::DIS::FAMILIES_APP}
\begin{tabular}{ccccccc}
\hline
Rank & Name & Min & Max & Mean & Std. dev.& \# \\
\hline
\hline
16 & Radon Coefficients (Fourier ()) & 0.002796 & 0.1572 & 0.07966 & 0.05538 &
12 \\
17 & Haralick Textures (Chebyshev (Wavelet ())) & 0.001528 & 0.2495 & 0.07426 &
0.08442 & 28 \\
18 & Haralick Textures (Fourier (Chebyshev ())) & 0.001438 & 0.2683 & 0.07406 &
0.08039 & 28 \\
19 & Haralick Textures (Chebyshev ()) & 0.0002557 & 0.2406 & 0.07387 & 0.08594
& 28 \\
20 & Tamura Textures (Wavelet (Fourier ())) & 0.008465 & 0.1561 & 0.07253 &
0.05688 & 6 \\
21 & Radon Coefficients (Wavelet (Fourier ())) & 0.0006693 & 0.1932 & 0.0725 &
0.06093 & 12 \\
22 & Radon Coefficients (Fourier (Edge ())) & 0.01066 & 0.1438 & 0.06954 &
0.04716 & 12 \\
23 & Fractal Features (Fourier (Edge ())) & 0.04659 & 0.07981 & 0.06902 &
0.01171 & 20 \\
24 & Radon Coefficients (Fourier (Wavelet ())) & 0.005142 & 0.1236 & 0.06871 &
0.04525 & 12 \\
25 & Pixel Intensity Statistics (Chebyshev ()) & 0.003185 & 0.1221 & 0.06781 &
0.05914 & 5 \\
26 & Tamura Textures (Wavelet (Edge ())) & 0.002864 & 0.1124 & 0.06567 &
0.04648 & 6 \\
27 & Gini Coefficient (Wavelet ()) & 0.06536 & 0.06536 & 0.06536 & -- & 1 \\
28 & Multiscale Histograms (Chebyshev (Fourier ())) & 2.761e-05 & 0.3119 &
0.06379 & 0.09224 & 24 \\
29 & Tamura Textures (Fourier (Wavelet ())) & 0.0 & 0.1739 & 0.06093 & 0.08504
& 6 \\
30 & Haralick Textures (Wavelet (Fourier ())) & 0.001761 & 0.2073 & 0.05406 &
0.05679 & 28 \\
31 & Edge Features () & 0.0 & 0.1329 & 0.05388 & 0.04086 & 28 \\
32 & Pixel Intensity Statistics (Edge ()) & 0.0 & 0.2079 & 0.05386 & 0.08666 &
5 \\
33 & Pixel Intensity Statistics (Fourier (Chebyshev ())) & 0.001726 & 0.1099 &
0.05127 & 0.04356 & 5 \\
34 & Zernike Coefficients (Fourier (Edge ())) & 0.002065 & 0.1793 & 0.0508 &
0.04392 & 72 \\
35 & Pixel Intensity Statistics (Wavelet (Edge ())) & 0.004129 & 0.1825 &
0.05042 & 0.07427 & 5 \\
36 & Haralick Textures (Chebyshev (Fourier ())) & 0.0007449 & 0.1662 & 0.04898
& 0.05044 & 28 \\
37 & Pixel Intensity Statistics (Wavelet (Fourier ())) & 0.009989 & 0.102 &
0.04794 & 0.03919 & 5 \\
38 & Pixel Intensity Statistics (Chebyshev (Fourier ())) & 0.0319 & 0.06915 &
0.0473 & 0.01647 & 5 \\
39 & Fractal Features (Chebyshev (Fourier ())) & 0.03638 & 0.05426 & 0.04568 &
0.005604 & 20 \\
40 & Fractal Features (Fourier (Chebyshev ())) & 0.004817 & 0.0715 & 0.04502 &
0.02301 & 20 \\
41 & Haralick Textures (Wavelet ()) & 0.0005351 & 0.3202 & 0.04411 & 0.07016 &
28 \\
42 & Radon Coefficients (Chebyshev (Wavelet ())) & 0.0004625 & 0.09457 &
0.04259 & 0.04378 & 12 \\
43 & Tamura Textures (Wavelet ()) & 0.002278 & 0.07851 & 0.04257 & 0.02963 & 6
\\
44 & Pixel Intensity Statistics (Chebyshev (Wavelet ())) & 0.0005042 & 0.0785 &
0.04232 & 0.03863 & 5 \\
45 & Chebyshev Coefficients (Fourier ()) & 0.0001485 & 0.1802 & 0.04178 &
0.05048 & 32 \\
46 & Gini Coefficient (Fourier (Chebyshev ())) & 0.04167 & 0.04167 & 0.04167 &
-- & 1 \\
47 & Haralick Textures (Wavelet (Edge ())) & 6.751e-05 & 0.2854 & 0.04003 &
0.06803 & 28 \\
48 & Pixel Intensity Statistics (Fourier (Edge ())) & 0.006477 & 0.07215 &
0.0386 & 0.03069 & 5 \\
49 & Chebyshev-Fourier Coefficients () & 0.00278 & 0.08729 & 0.03718 & 0.02502
& 32 \\
50 & Tamura Textures (Chebyshev (Fourier ())) & 0.0 & 0.1086 & 0.03678 &
0.03851 & 6 \\
51 & Haralick Textures (Fourier (Wavelet ())) & 0.003524 & 0.09116 & 0.03545 &
0.02924 & 28 \\
52 & Pixel Intensity Statistics (Fourier ()) & 0.006817 & 0.09874 & 0.03542 &
0.04006 & 5 \\
53 & Gabor Textures () & 0.005368 & 0.05098 & 0.03541 & 0.01482 & 7 \\
54 & Multiscale Histograms () & 0.0 & 0.06035 & 0.03538 & 0.01127 & 24 \\
55 & Pixel Intensity Statistics (Fourier (Wavelet ())) & 0.006814 & 0.08721 &
0.03479 & 0.03649 & 5 \\
56 & Chebyshev Coefficients (Fourier (Wavelet ())) & 7.333e-05 & 0.1373 &
0.03357 & 0.0354 & 32 \\
57 & Fractal Features (Wavelet (Edge ())) & 0.01615 & 0.08687 & 0.03158 &
0.01775 & 20 \\
58 & Haralick Textures (Fourier ()) & 0.005475 & 0.08353 & 0.03136 & 0.02556 &
28 \\
59 & Radon Coefficients (Chebyshev ()) & 0.001171 & 0.1105 & 0.03025 & 0.03528
& 12 \\
60 & Zernike Coefficients (Wavelet ()) & 0.001553 & 0.04834 & 0.02949 & 0.01235
& 72 \\
61 & Tamura Textures (Fourier (Chebyshev ())) & 0.0 & 0.05934 & 0.0291 &
0.02429 & 6 \\
62 & Tamura Textures (Fourier (Edge ())) & 0.0 & 0.07595 & 0.02848 & 0.03662 &
6 \\
63 & Comb Moments (Edge ()) & 0.003435 & 0.1302 & 0.02723 & 0.02894 & 48 \\
64 & Multiscale Histograms (Chebyshev ()) & 0.001093 & 0.08173 & 0.02662 &
0.02668 & 24 \\
65 & Multiscale Histograms (Chebyshev (Wavelet ())) & 0.00025 & 0.07265 &
0.02647 & 0.01974 & 24 \\
66 & Comb Moments (Fourier (Wavelet ())) & 0.0 & 0.1406 & 0.02633 & 0.039 & 48
\\
67 & Multiscale Histograms (Fourier (Wavelet ())) & 0.0 & 0.06156 & 0.02345 &
0.01592 & 24 \\
68 & Fractal Features (Edge ()) & 0.009364 & 0.06868 & 0.02333 & 0.01531 & 20 \\
69 & Zernike Coefficients () & 0.0004158 & 0.04813 & 0.02311 & 0.01263 & 72 \\
70 & Gini Coefficient (Fourier (Edge ())) & 0.02216 & 0.02216 & 0.02216 & -- &
1 \\

\end{tabular}
\end{table*}

\begin{table*}
\caption{Table \ref{TAB::DIS::FAMILIES} continued}
\begin{tabular}{ccccccc}
\hline
Rank & Name & Min & Max & Mean & Std. dev.& \# \\
\hline
\hline
71 & Tamura Textures (Edge ()) & 0.0006446 & 0.05768 & 0.02165 & 0.02062 & 6 \\
72 & Pixel Intensity Statistics (Wavelet ()) & 0.006814 & 0.05038 & 0.02161 &
0.01707 & 5 \\
73 & Chebyshev Coefficients () & 0.0008869 & 0.0634 & 0.02085 & 0.01768 & 32 \\
74 & Radon Coefficients (Chebyshev (Fourier ())) & 0.0008424 & 0.05026 &
0.02029 & 0.01923 & 12 \\
75 & Multiscale Histograms (Wavelet (Fourier ())) & 0.0002537 & 0.08552 &
0.02007 & 0.02656 & 24 \\
76 & Comb Moments () & 0.000969 & 0.0631 & 0.01988 & 0.01873 & 48 \\
77 & Multiscale Histograms (Edge ()) & 0.0 & 0.04351 & 0.01965 & 0.01608 & 24 \\
78 & Comb Moments (Chebyshev (Wavelet ())) & 2.809e-05 & 0.1204 & 0.01957 &
0.02995 & 48 \\
79 & Chebyshev Coefficients (Fourier (Edge ())) & 7.002e-05 & 0.1566 & 0.01886
& 0.03433 & 32 \\
80 & Comb Moments (Fourier (Edge ())) & 0.0 & 0.06139 & 0.0188 & 0.01759 & 48 \\
81 & Comb Moments (Chebyshev (Fourier ())) & 0.0001936 & 0.07083 & 0.01856 &
0.01756 & 48 \\
82 & Multiscale Histograms (Fourier ()) & 0.0 & 0.03878 & 0.01818 & 0.01083 &
24 \\
83 & Haralick Textures (Fourier (Edge ())) & 0.0011 & 0.0498 & 0.01772 &
0.01365 & 28 \\
84 & Chebyshev Coefficients (Edge ()) & 2.742e-05 & 0.05939 & 0.01769 & 0.01718
& 32 \\
85 & Pixel Intensity Statistics () & 0.0 & 0.05479 & 0.01738 & 0.02333 & 5 \\
86 & Multiscale Histograms (Wavelet ()) & 0.0 & 0.03136 & 0.01676 & 0.011 & 24
\\
87 & Tamura Textures () & 0.003395 & 0.03918 & 0.01556 & 0.01372 & 6 \\
88 & Comb Moments (Fourier ()) & 0.0 & 0.08391 & 0.01543 & 0.02206 & 48 \\
89 & Chebyshev-Fourier Coefficients (Edge ()) & 0.002144 & 0.03468 & 0.0149 &
0.008587 & 32 \\
90 & Otsu Object Features () & 2.923e-05 & 0.07006 & 0.01471 & 0.01543 & 34 \\
91 & Multiscale Histograms (Fourier (Edge ())) & 0.0 & 0.03262 & 0.01357 &
0.01125 & 24 \\
92 & Comb Moments (Fourier (Chebyshev ())) & 0.0002988 & 0.06055 & 0.01302 &
0.01223 & 48 \\
93 & Comb Moments (Wavelet (Edge ())) & 0.0005041 & 0.08158 & 0.01256 & 0.01493
& 48 \\
94 & Comb Moments (Wavelet ()) & 0.000263 & 0.05191 & 0.01214 & 0.01211 & 48 \\
95 & Chebyshev Coefficients (Wavelet ()) & 0.0006492 & 0.03937 & 0.01193 &
0.009978 & 32 \\
96 & Fractal Features (Fourier (Wavelet ())) & 0.006856 & 0.03016 & 0.01164 &
0.005313 & 20 \\
97 & Fractal Features () & 0.003373 & 0.02253 & 0.01111 & 0.006299 & 20 \\
98 & Comb Moments (Wavelet (Fourier ())) & 0.0 & 0.06691 & 0.01077 & 0.01314 &
48 \\
99 & Zernike Coefficients (Edge ()) & 0.0009013 & 0.03445 & 0.01066 & 0.006926
& 72 \\
100 & Fractal Features (Chebyshev (Wavelet ())) & 0.004038 & 0.02374 & 0.01037
& 0.004437 & 20 \\
101 & Fractal Features (Wavelet ()) & 0.001692 & 0.02562 & 0.01001 & 0.006148 &
20 \\
102 & Fractal Features (Wavelet (Fourier ())) & 0.00538 & 0.0117 & 0.009496 &
0.002046 & 20 \\
103 & Zernike Coefficients (Wavelet (Edge ())) & 0.0008492 & 0.02444 & 0.009097
& 0.005664 & 72 \\
104 & Gini Coefficient (Chebyshev (Fourier ())) & 0.00895 & 0.00895 & 0.00895 &
-- & 1 \\
105 & Radon Coefficients (Fourier (Chebyshev ())) & 0.002551 & 0.0178 &
0.008616 & 0.005505 & 12 \\
106 & Chebyshev Coefficients (Chebyshev ()) & 0.0002586 & 0.01941 & 0.008269 &
0.006129 & 32 \\
107 & Multiscale Histograms (Wavelet (Edge ())) & 0.0001181 & 0.02458 &
0.008168 & 0.009393 & 24 \\
108 & Fractal Features (Fourier ()) & 0.005944 & 0.02786 & 0.008154 & 0.004887
& 20 \\
109 & Multiscale Histograms (Fourier (Chebyshev ())) & 0.0 & 0.01491 & 0.00687
& 0.004261 & 24 \\
110 & Chebyshev-Fourier Coefficients (Fourier (Wavelet ())) & 0.0002502 &
0.05058 & 0.006706 & 0.01031 & 32 \\
111 & Chebyshev Coefficients (Wavelet (Edge ())) & 0.0005705 & 0.02501 &
0.006566 & 0.005566 & 32 \\
112 & Radon Coefficients (Wavelet (Edge ())) & 0.001386 & 0.0106 & 0.006331 &
0.002556 & 12 \\
113 & Fractal Features (Chebyshev ()) & 0.004724 & 0.007529 & 0.006328 &
0.0009132 & 20 \\
114 & Chebyshev-Fourier Coefficients (Fourier (Edge ())) & 7.099e-06 & 0.02821
& 0.005849 & 0.006827 & 32 \\
115 & Chebyshev-Fourier Coefficients (Wavelet (Edge ())) & 0.0004125 & 0.01329
& 0.005519 & 0.003601 & 32 \\
116 & Radon Coefficients (Edge ()) & 0.001494 & 0.007173 & 0.005212 & 0.001605
& 12 \\
117 & Comb Moments (Chebyshev ()) & 4.192e-05 & 0.04941 & 0.005006 & 0.01017 &
48 \\
118 & Radon Coefficients (Wavelet ()) & 0.0006035 & 0.008731 & 0.00455 &
0.002576 & 12 \\
119 & Radon Coefficients () & 0.0003923 & 0.01101 & 0.004297 & 0.003788 & 12 \\
120 & Chebyshev-Fourier Coefficients (Fourier ()) & 3.925e-05 & 0.01895 &
0.003691 & 0.004664 & 32 \\
121 & Chebyshev-Fourier Coefficients (Wavelet ()) & 0.0002408 & 0.01188 &
0.003343 & 0.002948 & 32 \\
122 & Chebyshev-Fourier Coefficients (Chebyshev ()) & 0.000334 & 0.004244 &
0.001783 & 0.001106 & 32 \\
123 & Inverse-Otsu Object Features () & 0.0 & 0.01933 & 0.001535 & 0.004358 &
34 \\
124 & Zernike Coefficients (Chebyshev ()) & 0.001355 & 0.001563 & 0.001451 &
5.332e-05 & 72 \\
\end{tabular}
\end{table*}
 \end{document}

%% file: journal_def.tex
\def\apj{ApJ}%
\def\apjl{ApJ}%
\def\apjs{ApJS}%
\def\ao{Appl.~Opt.}%
\def\aap{A\&A}%
\def\jcap{J. Cosmology Astropart. Phys.}%
\def\mnras{MNRAS}%
\def\prd{Phys.~Rev.~D}%
\def\nat{Nature}%
\def\physrep{Phys.~Rep.}%

%% file: mg_computer_vision.bbl
\begin{thebibliography}{}
\makeatletter
\relax
\def\mn@urlcharsother{\let\do\@makeother \do\$\do\&\do\#\do\^\do\_\do\%\do\~}
\def\mn@doi{\begingroup\mn@urlcharsother \@ifnextchar [ {\mn@doi@}
  {\mn@doi@[]}}
\def\mn@doi@[#1]#2{\def\@tempa{#1}\ifx\@tempa\@empty \href
  {http://dx.doi.org/#2} {doi:#2}\else \href {http://dx.doi.org/#2} {#1}\fi
  \endgroup}
\def\mn@eprint#1#2{\mn@eprint@#1:#2::\@nil}
\def\mn@eprint@arXiv#1{\href {http://arxiv.org/abs/#1} {{\tt arXiv:#1}}}
\def\mn@eprint@dblp#1{\href {http://dblp.uni-trier.de/rec/bibtex/#1.xml}
  {dblp:#1}}
\def\mn@eprint@#1:#2:#3:#4\@nil{\def\@tempa {#1}\def\@tempb {#2}\def\@tempc
  {#3}\ifx \@tempc \@empty \let \@tempc \@tempb \let \@tempb \@tempa \fi \ifx
  \@tempb \@empty \def\@tempb {arXiv}\fi \@ifundefined
  {mn@eprint@\@tempb}{\@tempb:\@tempc}{\expandafter \expandafter \csname
  mn@eprint@\@tempb\endcsname \expandafter{\@tempc}}}

\bibitem[\protect\citeauthoryear{{Abraham}, {van den Bergh}  \&
  {Nair}}{{Abraham} et~al.}{2003}]{Abraham2003}
{Abraham} R.~G.,  {van den Bergh} S.,   {Nair} P.,  2003, \mn@doi [\apj]
  {10.1086/373919}, \href {http://adsabs.harvard.edu/abs/2003ApJ...588..218A}
  {588, 218}

\bibitem[\protect\citeauthoryear{{Anderson} et~al.,}{{Anderson}
  et~al.}{2014}]{Anderson2014}
{Anderson} L.,  et~al., 2014, \mn@doi [\mnras] {10.1093/mnras/stu523}, \href
  {http://adsabs.harvard.edu/abs/2014MNRAS.441...24A} {441, 24}

\bibitem[\protect\citeauthoryear{{Barreira}, {Llinares}, {Bose}  \&
  {Li}}{{Barreira} et~al.}{2016}]{Barreira2016}
{Barreira} A.,  {Llinares} C.,  {Bose} S.,   {Li} B.,  2016, \mn@doi [\jcap]
  {10.1088/1475-7516/2016/05/001}, \href
  {http://adsabs.harvard.edu/abs/2016JCAP...05..001B} {5, 001}

\bibitem[\protect\citeauthoryear{{Bartelmann}}{{Bartelmann}}{2010}]{Bartelmann2010a}
{Bartelmann} M.,  2010, \mn@doi [Classical and Quantum Gravity]
  {10.1088/0264-9381/27/23/233001}, \href
  {http://adsabs.harvard.edu/abs/2010CQGra..27w3001B} {27, 233001}

\bibitem[\protect\citeauthoryear{{Bartelmann} \& {Schneider}}{{Bartelmann} \&
  {Schneider}}{2001}]{Bartelmann2001}
{Bartelmann} M.,  {Schneider} P.,  2001, \physrep, \href
  {http://cdsads.u-strasbg.fr/abs/2001PhR...340..291B} {340, 291}

\bibitem[\protect\citeauthoryear{{Becker} \& {Kravtsov}}{{Becker} \&
  {Kravtsov}}{2011}]{Becker2011}
{Becker} M.~R.,  {Kravtsov} A.~V.,  2011, \mn@doi [\apj]
  {10.1088/0004-637X/740/1/25}, \href
  {http://adsabs.harvard.edu/abs/2011ApJ...740...25B} {740, 25}

\bibitem[\protect\citeauthoryear{{Behroozi}, {Wechsler}  \& {Wu}}{{Behroozi}
  et~al.}{2013}]{Behroozi2013}
{Behroozi} P.~S.,  {Wechsler} R.~H.,   {Wu} H.-Y.,  2013, \mn@doi [\apj]
  {10.1088/0004-637X/762/2/109}, \href
  {http://adsabs.harvard.edu/abs/2013ApJ...762..109B} {762, 109}

\bibitem[\protect\citeauthoryear{Bengtsson \& Rodenacker}{Bengtsson \&
  Rodenacker}{2003}]{Bengtsson2003}
Bengtsson E.,  Rodenacker K.,  2003, Analytical Cellular Pathology, 24, 1

\bibitem[\protect\citeauthoryear{{Betoule} et~al.,}{{Betoule}
  et~al.}{2014}]{Betoule2014}
{Betoule} M.,  et~al., 2014, \mn@doi [\aap] {10.1051/0004-6361/201423413},
  \href {http://adsabs.harvard.edu/abs/2014A%26A...568A..22B} {568, A22}

\bibitem[\protect\citeauthoryear{Bishop}{Bishop}{2006}]{Bishop2006a}
Bishop C.~M.,  2006, Pattern Recognition and Machine Learning (Information
  Science and Statistics).
Springer-Verlag New York, Inc., Secaucus, NJ, USA

\bibitem[\protect\citeauthoryear{{Boylan-Kolchin}, {Bullock}  \&
  {Kaplinghat}}{{Boylan-Kolchin} et~al.}{2012}]{Boylan-Kolchin2012}
{Boylan-Kolchin} M.,  {Bullock} J.~S.,   {Kaplinghat} M.,  2012, \mn@doi
  [\mnras] {10.1111/j.1365-2966.2012.20695.x}, \href
  {http://adsabs.harvard.edu/abs/2012MNRAS.422.1203B} {422, 1203}

\bibitem[\protect\citeauthoryear{{Bozek}, {Boylan-Kolchin}, {Horiuchi},
  {Garrison-Kimmel}, {Abazajian}  \& {Bullock}}{{Bozek}
  et~al.}{2016}]{Bozek2016}
{Bozek} B.,  {Boylan-Kolchin} M.,  {Horiuchi} S.,  {Garrison-Kimmel} S.,
  {Abazajian} K.,   {Bullock} J.~S.,  2016, \mn@doi [\mnras]
  {10.1093/mnras/stw688}, \href
  {http://adsabs.harvard.edu/abs/2016MNRAS.459.1489B} {459, 1489}

\bibitem[\protect\citeauthoryear{{Brada{\v c}} et~al.,}{{Brada{\v c}}
  et~al.}{2009}]{Bradav2009}
{Brada{\v c}} M.,  et~al., 2009, \mn@doi [\apj] {10.1088/0004-637X/706/2/1201},
  \href {http://adsabs.harvard.edu/abs/2009ApJ...706.1201B} {706, 1201}

\bibitem[\protect\citeauthoryear{{Caminha} et~al.,}{{Caminha}
  et~al.}{2016}]{Caminha2016}
{Caminha} G.~B.,  et~al., 2016, \mn@doi [\aap] {10.1051/0004-6361/201527670},
  \href {http://adsabs.harvard.edu/abs/2016A%26A...587A..80C} {587, A80}

\bibitem[\protect\citeauthoryear{{Clifton}, {Ferreira}, {Padilla}  \&
  {Skordis}}{{Clifton} et~al.}{2012}]{Clifton2012}
{Clifton} T.,  {Ferreira} P.~G.,  {Padilla} A.,   {Skordis} C.,  2012, \mn@doi
  [\physrep] {10.1016/j.physrep.2012.01.001}, \href
  {http://adsabs.harvard.edu/abs/2012PhR...513....1C} {513, 1}

\bibitem[\protect\citeauthoryear{{Clowe}, {Brada{\v c}}, {Gonzalez},
  {Markevitch}, {Randall}, {Jones}  \& {Zaritsky}}{{Clowe}
  et~al.}{2006}]{Clowe2006}
{Clowe} D.,  {Brada{\v c}} M.,  {Gonzalez} A.~H.,  {Markevitch} M.,  {Randall}
  S.~W.,  {Jones} C.,   {Zaritsky} D.,  2006, \mn@doi [\apjl] {10.1086/508162},
  \href {http://cdsads.u-strasbg.fr/abs/2006ApJ...648L.109C} {648, L109}

\bibitem[\protect\citeauthoryear{{Dawson} et~al.,}{{Dawson}
  et~al.}{2012}]{Dawson2012}
{Dawson} W.~A.,  et~al., 2012, \mn@doi [\apjl] {10.1088/2041-8205/747/2/L42},
  \href {http://adsabs.harvard.edu/abs/2012ApJ...747L..42D} {747, L42}

\bibitem[\protect\citeauthoryear{{Diego} et~al.,}{{Diego}
  et~al.}{2015}]{Diego2015}
{Diego} J.~M.,  et~al., 2015, \mn@doi [\mnras] {10.1093/mnras/stu2064}, \href
  {http://adsabs.harvard.edu/abs/2015MNRAS.446..683D} {446, 683}

\bibitem[\protect\citeauthoryear{{Diego}, {Broadhurst}, {Wong}, {Silk}, {Lim},
  {Zheng}, {Lam}  \& {Ford}}{{Diego} et~al.}{2016}]{Diego2016}
{Diego} J.~M.,  {Broadhurst} T.,  {Wong} J.,  {Silk} J.,  {Lim} J.,  {Zheng}
  W.,  {Lam} D.,   {Ford} H.,  2016, \mn@doi [\mnras] {10.1093/mnras/stw865},
  \href {http://adsabs.harvard.edu/abs/2016MNRAS.459.3447D} {459, 3447}

\bibitem[\protect\citeauthoryear{{Dubois} et~al.,}{{Dubois}
  et~al.}{2014}]{Dubois2014}
{Dubois} Y.,  et~al., 2014, \mn@doi [\mnras] {10.1093/mnras/stu1227}, \href
  {http://adsabs.harvard.edu/abs/2014MNRAS.444.1453D} {444, 1453}

\bibitem[\protect\citeauthoryear{{Dvali}, {Gabadadze}  \& {Porrati}}{{Dvali}
  et~al.}{2000}]{Dvali2000}
{Dvali} G.,  {Gabadadze} G.,   {Porrati} M.,  2000, \mn@doi [Physics Letters B]
  {10.1016/S0370-2693(00)00669-9}, \href
  {http://adsabs.harvard.edu/abs/2000PhLB..485..208D} {485, 208}

\bibitem[\protect\citeauthoryear{{Florian}, {Gladders}, {Li}  \&
  {Sharon}}{{Florian} et~al.}{2016}]{Florian2016}
{Florian} M.~K.,  {Gladders} M.~D.,  {Li} N.,   {Sharon} K.,  2016, \mn@doi
  [\apjl] {10.3847/2041-8205/816/2/L23}, \href
  {http://adsabs.harvard.edu/abs/2016ApJ...816L..23F} {816, L23}

\bibitem[\protect\citeauthoryear{Fogel \& Sagi}{Fogel \&
  Sagi}{1989}]{Fogel1989}
Fogel I.,  Sagi D.,  1989, \mn@doi [Biological Cybernetics]
  {10.1007/BF00204594}, 61, 103

\bibitem[\protect\citeauthoryear{{Gonzalez} \& {Woods}}{{Gonzalez} \&
  {Woods}}{2007}]{Gonzalez2007}
{Gonzalez} R.~C.,  {Woods} R.~E.,  2007, Digital Image Processing, 3 edn.
Prentice Hall

\bibitem[\protect\citeauthoryear{Grigorescu, Petkov  \& Kruizinga}{Grigorescu
  et~al.}{2002}]{Grigorescu2002}
Grigorescu S.~E.,  Petkov N.,   Kruizinga P.,  2002, \mn@doi [IEEE Transactions
  on Image Processing] {10.1109/TIP.2002.804262}, 11, 1160

\bibitem[\protect\citeauthoryear{Haralick, Shanmugam  \& Dinstein}{Haralick
  et~al.}{1973}]{Haralick1973}
Haralick R.~M.,  Shanmugam K.,   Dinstein I.,  1973, \mn@doi [IEEE Transactions
  on Systems, Man, and Cybernetics] {10.1109/TSMC.1973.4309314}, SMC-3, 610

\bibitem[\protect\citeauthoryear{{Harvey}, {Massey}, {Kitching}, {Taylor}  \&
  {Tittley}}{{Harvey} et~al.}{2015}]{Harvey2015}
{Harvey} D.,  {Massey} R.,  {Kitching} T.,  {Taylor} A.,   {Tittley} E.,  2015,
  \mn@doi [Science] {10.1126/science.1261381}, \href
  {http://adsabs.harvard.edu/abs/2015Sci...347.1462H} {347, 1462}

\bibitem[\protect\citeauthoryear{{Heitmann} et~al.,}{{Heitmann}
  et~al.}{2016}]{Heitmann2016}
{Heitmann} K.,  et~al., 2016, \mn@doi [\apj] {10.3847/0004-637X/820/2/108},
  \href {http://adsabs.harvard.edu/abs/2016ApJ...820..108H} {820, 108}

\bibitem[\protect\citeauthoryear{Heller \& Ghahramani}{Heller \&
  Ghahramani}{2006}]{Heller2006}
Heller K.~A.,  Ghahramani Z.,  2006, in 2006 IEEE Computer Society Conference
  on Computer Vision and Pattern Recognition (CVPR'06). pp 2110--2117,
  \mn@doi{10.1109/CVPR.2006.41}

\bibitem[\protect\citeauthoryear{{Hildebrandt} et~al.,}{{Hildebrandt}
  et~al.}{2017}]{Hildebrandt2017}
{Hildebrandt} H.,  et~al., 2017, \mn@doi [\mnras] {10.1093/mnras/stw2805},
  \href {http://adsabs.harvard.edu/abs/2017MNRAS.465.1454H} {465, 1454}

\bibitem[\protect\citeauthoryear{{Hinterbichler} \& {Khoury}}{{Hinterbichler}
  \& {Khoury}}{2010}]{Hinterbichler2010}
{Hinterbichler} K.,  {Khoury} J.,  2010, \mn@doi [Physical Review Letters]
  {10.1103/PhysRevLett.104.231301}, \href
  {http://adsabs.harvard.edu/abs/2010PhRvL.104w1301H} {104, 231301}

\bibitem[\protect\citeauthoryear{{Hoekstra}, {Hartlap}, {Hilbert}  \& {van
  Uitert}}{{Hoekstra} et~al.}{2011}]{Hoekstra2011}
{Hoekstra} H.,  {Hartlap} J.,  {Hilbert} S.,   {van Uitert} E.,  2011, \mn@doi
  [\mnras] {10.1111/j.1365-2966.2010.18053.x}, \href
  {http://adsabs.harvard.edu/abs/2011MNRAS.412.2095H} {412, 2095}

\bibitem[\protect\citeauthoryear{{Hopkins}, {Kere{\v s}}, {O{\~n}orbe},
  {Faucher-Gigu{\`e}re}, {Quataert}, {Murray}  \& {Bullock}}{{Hopkins}
  et~al.}{2014}]{Hopkins2014}
{Hopkins} P.~F.,  {Kere{\v s}} D.,  {O{\~n}orbe} J.,  {Faucher-Gigu{\`e}re}
  C.-A.,  {Quataert} E.,  {Murray} N.,   {Bullock} J.~S.,  2014, \mn@doi
  [\mnras] {10.1093/mnras/stu1738}, \href
  {http://adsabs.harvard.edu/abs/2014MNRAS.445..581H} {445, 581}

\bibitem[\protect\citeauthoryear{{Hu} \& {Sawicki}}{{Hu} \&
  {Sawicki}}{2007}]{Hu2007}
{Hu} W.,  {Sawicki} I.,  2007, \mn@doi [\prd] {10.1103/PhysRevD.76.064004},
  \href {http://adsabs.harvard.edu/abs/2007PhRvD..76f4004H} {76, 064004}

\bibitem[\protect\citeauthoryear{{Jauzac} et~al.,}{{Jauzac}
  et~al.}{2015}]{Jauzac2015}
{Jauzac} M.,  et~al., 2015, \mn@doi [\mnras] {10.1093/mnras/stu2425}, \href
  {http://adsabs.harvard.edu/abs/2015MNRAS.446.4132J} {446, 4132}

\bibitem[\protect\citeauthoryear{{Jee}, {Hughes}, {Menanteau}, {Sif{\'o}n},
  {Mandelbaum}, {Barrientos}, {Infante}  \& {Ng}}{{Jee}
  et~al.}{2014}]{Jee2014a}
{Jee} M.~J.,  {Hughes} J.~P.,  {Menanteau} F.,  {Sif{\'o}n} C.,  {Mandelbaum}
  R.,  {Barrientos} L.~F.,  {Infante} L.,   {Ng} K.~Y.,  2014, \mn@doi [\apj]
  {10.1088/0004-637X/785/1/20}, \href
  {http://adsabs.harvard.edu/abs/2014ApJ...785...20J} {785, 20}

\bibitem[\protect\citeauthoryear{{Johnson}, {Sharon}, {Bayliss}, {Gladders},
  {Coe}  \& {Ebeling}}{{Johnson} et~al.}{2014}]{Johnson2014a}
{Johnson} T.~L.,  {Sharon} K.,  {Bayliss} M.~B.,  {Gladders} M.~D.,  {Coe} D.,
   {Ebeling} H.,  2014, \mn@doi [\apj] {10.1088/0004-637X/797/1/48}, \href
  {http://adsabs.harvard.edu/abs/2014ApJ...797...48J} {797, 48}

\bibitem[\protect\citeauthoryear{{Jullo} \& {Kneib}}{{Jullo} \&
  {Kneib}}{2009}]{Jullo2009}
{Jullo} E.,  {Kneib} J.,  2009, \mn@doi [\mnras]
  {10.1111/j.1365-2966.2009.14654.x}, \href
  {http://adsabs.harvard.edu/abs/2009MNRAS.395.1319J} {395, 1319}

\bibitem[\protect\citeauthoryear{{Kauffmann}, {White}  \&
  {Guiderdoni}}{{Kauffmann} et~al.}{1993}]{Kauffmann1993}
{Kauffmann} G.,  {White} S.~D.~M.,   {Guiderdoni} B.,  1993, \mn@doi [\mnras]
  {10.1093/mnras/264.1.201}, \href
  {http://adsabs.harvard.edu/abs/1993MNRAS.264..201K} {264, 201}

\bibitem[\protect\citeauthoryear{{Kimm}, {Cen}, {Devriendt}, {Dubois}  \&
  {Slyz}}{{Kimm} et~al.}{2015}]{Kimm2015}
{Kimm} T.,  {Cen} R.,  {Devriendt} J.,  {Dubois} Y.,   {Slyz} A.,  2015,
  \mn@doi [\mnras] {10.1093/mnras/stv1211}, \href
  {http://adsabs.harvard.edu/abs/2015MNRAS.451.2900K} {451, 2900}

\bibitem[\protect\citeauthoryear{{L'Huillier}, {Winther}, {Mota}, {Park}  \&
  {Kim}}{{L'Huillier} et~al.}{2017}]{LHuillier2017}
{L'Huillier} B.,  {Winther} H.~A.,  {Mota} D.~F.,  {Park} C.,   {Kim} J.,
  2017, preprint, \href {http://adsabs.harvard.edu/abs/2017arXiv170307357L} {}
  (\mn@eprint {arXiv} {1703.07357})

\bibitem[\protect\citeauthoryear{{Llinares}, {Mota}  \& {Winther}}{{Llinares}
  et~al.}{2014}]{Llinares2014}
{Llinares} C.,  {Mota} D.~F.,   {Winther} H.~A.,  2014, \mn@doi [\aap]
  {10.1051/0004-6361/201322412}, \href
  {http://adsabs.harvard.edu/abs/2014A%26A...562A..78L} {562, A78}

\bibitem[\protect\citeauthoryear{{Lovell}, {Frenk}, {Eke}, {Jenkins}, {Gao}  \&
  {Theuns}}{{Lovell} et~al.}{2014}]{Lovell2014}
{Lovell} M.~R.,  {Frenk} C.~S.,  {Eke} V.~R.,  {Jenkins} A.,  {Gao} L.,
  {Theuns} T.,  2014, \mn@doi [\mnras] {10.1093/mnras/stt2431}, \href
  {http://adsabs.harvard.edu/abs/2014MNRAS.439..300L} {439, 300}

\bibitem[\protect\citeauthoryear{{Massey} et~al.,}{{Massey}
  et~al.}{2007}]{Massey2007}
{Massey} R.,  et~al., 2007, \mn@doi [\nat] {10.1038/nature05497}, \href
  {http://adsabs.harvard.edu/abs/2007Natur.445..286M} {445, 286}

\bibitem[\protect\citeauthoryear{{Massey} et~al.,}{{Massey}
  et~al.}{2015}]{Massey2015}
{Massey} R.,  et~al., 2015, \mn@doi [\mnras] {10.1093/mnras/stv467}, \href
  {http://adsabs.harvard.edu/abs/2015MNRAS.449.3393M} {449, 3393}

\bibitem[\protect\citeauthoryear{{McCarthy}, {Schaye}, {Bird}  \& {Le
  Brun}}{{McCarthy} et~al.}{2017}]{McCarthy2017}
{McCarthy} I.~G.,  {Schaye} J.,  {Bird} S.,   {Le Brun} A.~M.~C.,  2017,
  \mn@doi [\mnras] {10.1093/mnras/stw2792}, \href
  {http://adsabs.harvard.edu/abs/2017MNRAS.465.2936M} {465, 2936}

\bibitem[\protect\citeauthoryear{{McGaugh}, {Lelli}  \& {Schombert}}{{McGaugh}
  et~al.}{2016}]{McGaugh2016}
{McGaugh} S.~S.,  {Lelli} F.,   {Schombert} J.~M.,  2016, \mn@doi [Physical
  Review Letters] {10.1103/PhysRevLett.117.201101}, \href
  {http://adsabs.harvard.edu/abs/2016PhRvL.117t1101M} {117, 201101}

\bibitem[\protect\citeauthoryear{{Mead}, {Heymans}, {Lombriser}, {Peacock},
  {Steele}  \& {Winther}}{{Mead} et~al.}{2016}]{Mead2016}
{Mead} A.~J.,  {Heymans} C.,  {Lombriser} L.,  {Peacock} J.~A.,  {Steele}
  O.~I.,   {Winther} H.~A.,  2016, \mn@doi [\mnras] {10.1093/mnras/stw681},
  \href {http://adsabs.harvard.edu/abs/2016MNRAS.459.1468M} {459, 1468}

\bibitem[\protect\citeauthoryear{{Medezinski}, {Umetsu}, {Okabe}, {Nonino},
  {Molnar}, {Massey}, {Dupke}  \& {Merten}}{{Medezinski}
  et~al.}{2016}]{Medezinski2016}
{Medezinski} E.,  {Umetsu} K.,  {Okabe} N.,  {Nonino} M.,  {Molnar} S.,
  {Massey} R.,  {Dupke} R.,   {Merten} J.,  2016, \mn@doi [\apj]
  {10.3847/0004-637X/817/1/24}, \href
  {http://adsabs.harvard.edu/abs/2016ApJ...817...24M} {817, 24}

\bibitem[\protect\citeauthoryear{{Melchior} et~al.,}{{Melchior}
  et~al.}{2015}]{Melchior2015}
{Melchior} P.,  et~al., 2015, \mn@doi [\mnras] {10.1093/mnras/stv398}, \href
  {http://adsabs.harvard.edu/abs/2015MNRAS.449.2219M} {449, 2219}

\bibitem[\protect\citeauthoryear{{Meneghetti}, {Rasia}, {Merten}, {Bellagamba},
  {Ettori}, {Mazzotta}, {Dolag}  \& {Marri}}{{Meneghetti}
  et~al.}{2010}]{Meneghetti2010a}
{Meneghetti} M.,  {Rasia} E.,  {Merten} J.,  {Bellagamba} F.,  {Ettori} S.,
  {Mazzotta} P.,  {Dolag} K.,   {Marri} S.,  2010, \mn@doi [\aap]
  {10.1051/0004-6361/200913222}, \href
  {http://adsabs.harvard.edu/abs/2010A%26A...514A..93M} {514, A93+}

\bibitem[\protect\citeauthoryear{{Meneghetti} et~al.,}{{Meneghetti}
  et~al.}{2016}]{Meneghetti2016}
{Meneghetti} M.,  et~al., 2016, preprint, \href
  {http://adsabs.harvard.edu/abs/2016arXiv160604548M} {} (\mn@eprint {arXiv}
  {1606.04548})

\bibitem[\protect\citeauthoryear{{Merten}}{{Merten}}{2016}]{Merten2016}
{Merten} J.,  2016, \mn@doi [\mnras] {10.1093/mnras/stw1413}, \href
  {http://adsabs.harvard.edu/abs/2016MNRAS.461.2328M} {461, 2328}

\bibitem[\protect\citeauthoryear{{Merten}, {Cacciato}, {Meneghetti}, {Mignone}
  \& {Bartelmann}}{{Merten} et~al.}{2009}]{Merten2009}
{Merten} J.,  {Cacciato} M.,  {Meneghetti} M.,  {Mignone} C.,   {Bartelmann}
  M.,  2009, \mn@doi [\aap] {10.1051/0004-6361/200810372}, \href
  {http://adsabs.harvard.edu/abs/2009A%26A...500..681M} {500, 681}

\bibitem[\protect\citeauthoryear{{Merten} et~al.,}{{Merten}
  et~al.}{2011}]{Merten2011}
{Merten} J.,  et~al., 2011, \mn@doi [\mnras]
  {10.1111/j.1365-2966.2011.19266.x}, \href
  {http://adsabs.harvard.edu/abs/2011MNRAS.417..333M} {417, 333}

\bibitem[\protect\citeauthoryear{{Merten} et~al.,}{{Merten}
  et~al.}{2015}]{Merten2015}
{Merten} J.,  et~al., 2015, \mn@doi [\apj] {10.1088/0004-637X/806/1/4}, \href
  {http://adsabs.harvard.edu/abs/2015ApJ...806....4M} {806, 4}

\bibitem[\protect\citeauthoryear{{Nelson}, {Genel}, {Pillepich},
  {Vogelsberger}, {Springel}  \& {Hernquist}}{{Nelson}
  et~al.}{2016}]{Nelson2016}
{Nelson} D.,  {Genel} S.,  {Pillepich} A.,  {Vogelsberger} M.,  {Springel} V.,
   {Hernquist} L.,  2016, \mn@doi [\mnras] {10.1093/mnras/stw1191}, \href
  {http://adsabs.harvard.edu/abs/2016MNRAS.460.2881N} {460, 2881}

\bibitem[\protect\citeauthoryear{{Newman}, {Treu}, {Ellis}, {Sand}, {Nipoti},
  {Richard}  \& {Jullo}}{{Newman} et~al.}{2013}]{Newman2013}
{Newman} A.~B.,  {Treu} T.,  {Ellis} R.~S.,  {Sand} D.~J.,  {Nipoti} C.,
  {Richard} J.,   {Jullo} E.,  2013, \mn@doi [\apj]
  {10.1088/0004-637X/765/1/24}, \href
  {http://adsabs.harvard.edu/abs/2013ApJ...765...24N} {765, 24}

\bibitem[\protect\citeauthoryear{{Oman} et~al.,}{{Oman}
  et~al.}{2015}]{Oman2015}
{Oman} K.~A.,  et~al., 2015, \mn@doi [\mnras] {10.1093/mnras/stv1504}, \href
  {http://adsabs.harvard.edu/abs/2015MNRAS.452.3650O} {452, 3650}

\bibitem[\protect\citeauthoryear{Orlov, Johnston, Macura, Wolkow  \&
  Goldberg}{Orlov et~al.}{2006}]{Orlov2006}
Orlov N.,  Johnston J.,  Macura T.,  Wolkow C.,   Goldberg I.,  2006, in 3rd
  IEEE International Symposium on Biomedical Imaging: Nano to Macro, 2006.. pp
  1152--1155, \mn@doi{10.1109/ISBI.2006.1625127}

\bibitem[\protect\citeauthoryear{Orlov, Shamir, Macura, Johnston, Eckley  \&
  Goldberg}{Orlov et~al.}{2008}]{Orlov2008}
Orlov N.,  Shamir L.,  Macura T.,  Johnston J.,  Eckley D.~M.,   Goldberg
  I.~G.,  2008, Pattern recognition letters, 29, 1684

\bibitem[\protect\citeauthoryear{Otsu}{Otsu}{1979}]{Otsu1979}
Otsu N.,  1979, \mn@doi [IEEE Transactions on Systems, Man, and Cybernetics]
  {10.1109/TSMC.1979.4310076}, 9, 62

\bibitem[\protect\citeauthoryear{{Peter}, {Rocha}, {Bullock}  \&
  {Kaplinghat}}{{Peter} et~al.}{2013}]{Peter2013}
{Peter} A.~H.~G.,  {Rocha} M.,  {Bullock} J.~S.,   {Kaplinghat} M.,  2013,
  \mn@doi [\mnras] {10.1093/mnras/sts535}, \href
  {http://adsabs.harvard.edu/abs/2013MNRAS.tmp..600P} {p.~600}

\bibitem[\protect\citeauthoryear{{Planck Collaboration} et~al.,}{{Planck
  Collaboration} et~al.}{2016a}]{PlanckCollaboration2016}
{Planck Collaboration} et~al., 2016a, \mn@doi [\aap]
  {10.1051/0004-6361/201525830}, \href
  {http://adsabs.harvard.edu/abs/2016A%26A...594A..13P} {594, A13}

\bibitem[\protect\citeauthoryear{{Planck Collaboration} et~al.,}{{Planck
  Collaboration} et~al.}{2016b}]{PlanckCollaboration2016a}
{Planck Collaboration} et~al., 2016b, \mn@doi [\aap]
  {10.1051/0004-6361/201525833}, \href
  {http://adsabs.harvard.edu/abs/2016A%26A...594A..24P} {594, A24}

\bibitem[\protect\citeauthoryear{Prewitt}{Prewitt}{1970}]{Prewitt1970a}
Prewitt J.,  1970, Picture Processing and Psychopictorics.
New York: Academic Press

\bibitem[\protect\citeauthoryear{Radon}{Radon}{1917}]{Radon1917}
Radon J.,  1917, in Berichte \"{u}ber die Verhandlungen der
  K\"{o}niglich-S\"{a}chsischen Akademie der Wissenschaften zu Leipzig.
  Teubner, pp 262--277, \mn@doi{doi:10.1109/TMI.1986.4307775}, \url
  {https://dx.doi.org/10.1109%2FTMI.1986.4307775}

\bibitem[\protect\citeauthoryear{{Read}, {Agertz}  \& {Collins}}{{Read}
  et~al.}{2016}]{Read2016}
{Read} J.~I.,  {Agertz} O.,   {Collins} M.~L.~M.,  2016, \mn@doi [\mnras]
  {10.1093/mnras/stw713}, \href
  {http://adsabs.harvard.edu/abs/2016MNRAS.459.2573R} {459, 2573}

\bibitem[\protect\citeauthoryear{{Rocha}, {Peter}, {Bullock}, {Kaplinghat},
  {Garrison-Kimmel}, {O{\~n}orbe}  \& {Moustakas}}{{Rocha}
  et~al.}{2013}]{Rocha2013}
{Rocha} M.,  {Peter} A.~H.~G.,  {Bullock} J.~S.,  {Kaplinghat} M.,
  {Garrison-Kimmel} S.,  {O{\~n}orbe} J.,   {Moustakas} L.~A.,  2013, \mn@doi
  [\mnras] {10.1093/mnras/sts514}, \href
  {http://adsabs.harvard.edu/abs/2013MNRAS.tmp..560R} {p.~560}

\bibitem[\protect\citeauthoryear{{Schaye} et~al.,}{{Schaye}
  et~al.}{2015}]{Schaye2015}
{Schaye} J.,  et~al., 2015, \mn@doi [\mnras] {10.1093/mnras/stu2058}, \href
  {http://adsabs.harvard.edu/abs/2015MNRAS.446..521S} {446, 521}

\bibitem[\protect\citeauthoryear{{Schutter} \& {Shamir}}{{Schutter} \&
  {Shamir}}{2015}]{Schutter2015a}
{Schutter} A.,  {Shamir} L.,  2015, \mn@doi [Astronomy and Computing]
  {10.1016/j.ascom.2015.05.002}, \href
  {http://adsabs.harvard.edu/abs/2015A%26C....12...60S} {12, 60}

\bibitem[\protect\citeauthoryear{{Schwinn}, {Jauzac}, {Baugh}, {Bartelmann},
  {Eckert}, {Harvey}, {Natarajan}  \& {Massey}}{{Schwinn}
  et~al.}{2017}]{Schwinn2017}
{Schwinn} J.,  {Jauzac} M.,  {Baugh} C.~M.,  {Bartelmann} M.,  {Eckert} D.,
  {Harvey} D.,  {Natarajan} P.,   {Massey} R.,  2017, \mn@doi [\mnras]
  {10.1093/mnras/stx277}, \href
  {http://adsabs.harvard.edu/abs/2017MNRAS.467.2913S} {467, 2913}

\bibitem[\protect\citeauthoryear{Shamir, Orlov, Eckley, Macura, Johnston  \&
  Goldberg}{Shamir et~al.}{2008}]{Shamir2008}
Shamir L.,  Orlov N.,  Eckley D.~M.,  Macura T.,  Johnston J.,   Goldberg I.,
  2008, \mn@doi [Source Code for Biology and Medicine]
  {10.1186/1751-0473-3-13}, 3, 13

\bibitem[\protect\citeauthoryear{{Shamir}, {Delaney}, {Orlov}, {Eckley}  \&
  {Goldberg}}{{Shamir} et~al.}{2010}]{Shamir2010}
{Shamir} L.,  {Delaney} J.,  {Orlov} N.,  {Eckley} D.,   {Goldberg} I.,  2010,
  \mn@doi [PLoS Comput Biol 6(11): e1000974] {doi:10.1371/journal.pcbi.1000974}

\bibitem[\protect\citeauthoryear{Tamura, Mori  \& Yamawaki}{Tamura
  et~al.}{1978}]{Tamura1978}
Tamura H.,  Mori S.,   Yamawaki T.,  1978, \mn@doi [IEEE Transactions on
  Systems, Man, and Cybernetics] {10.1109/TSMC.1978.4309999}, 8, 460

\bibitem[\protect\citeauthoryear{Teague}{Teague}{1980}]{Teague1980}
Teague M.~R.,  1980, \mn@doi [J. Opt. Soc. Am.] {10.1364/JOSA.70.000920}, 70,
  920

\bibitem[\protect\citeauthoryear{{Teyssier}}{{Teyssier}}{2002}]{Teyssier2002}
{Teyssier} R.,  2002, \mn@doi [\aap] {10.1051/0004-6361:20011817}, \href
  {http://adsabs.harvard.edu/abs/2002A%26A...385..337T} {385, 337}

\bibitem[\protect\citeauthoryear{{Treu} et~al.,}{{Treu}
  et~al.}{2016}]{Treu2016}
{Treu} T.,  et~al., 2016, \mn@doi [\apj] {10.3847/0004-637X/817/1/60}, \href
  {http://adsabs.harvard.edu/abs/2016ApJ...817...60T} {817, 60}

\bibitem[\protect\citeauthoryear{{Umetsu} et~al.,}{{Umetsu}
  et~al.}{2014}]{Umetsu2014}
{Umetsu} K.,  et~al., 2014, \mn@doi [\apj] {10.1088/0004-637X/795/2/163}, \href
  {http://adsabs.harvard.edu/abs/2014ApJ...795..163U} {795, 163}

\bibitem[\protect\citeauthoryear{{Vogelsberger}, {Zavala}, {Simpson}  \&
  {Jenkins}}{{Vogelsberger} et~al.}{2014}]{Vogelsberger2014}
{Vogelsberger} M.,  {Zavala} J.,  {Simpson} C.,   {Jenkins} A.,  2014, \mn@doi
  [\mnras] {10.1093/mnras/stu1713}, \href
  {http://adsabs.harvard.edu/abs/2014MNRAS.444.3684V} {444, 3684}

\bibitem[\protect\citeauthoryear{{Walker} \& {Pe{\~n}arrubia}}{{Walker} \&
  {Pe{\~n}arrubia}}{2011}]{Walker2011}
{Walker} M.~G.,  {Pe{\~n}arrubia} J.,  2011, \mn@doi [\apj]
  {10.1088/0004-637X/742/1/20}, \href
  {http://adsabs.harvard.edu/abs/2011ApJ...742...20W} {742, 20}

\bibitem[\protect\citeauthoryear{{Winther} et~al.,}{{Winther}
  et~al.}{2015}]{Winther2015a}
{Winther} H.~A.,  et~al., 2015, \mn@doi [\mnras] {10.1093/mnras/stv2253}, \href
  {http://adsabs.harvard.edu/abs/2015MNRAS.454.4208W} {454, 4208}

\bibitem[\protect\citeauthoryear{Wu, Chen  \& Hsieh}{Wu et~al.}{1992}]{Wu1992}
Wu C.-M.,  Chen Y.-C.,   Hsieh K.-S.,  1992, \mn@doi [IEEE Transactions on
  Medical Imaging] {10.1109/42.141636}, 11, 141

\bibitem[\protect\citeauthoryear{Yavlinsky, Heesch  \& R{\"u}ger}{Yavlinsky
  et~al.}{2006}]{Yavlinsky2006}
Yavlinsky A.,  Heesch D.,   R{\"u}ger S.,  2006, in Image and Video Retrieval.
  pp 537--540, \url {http://oro.open.ac.uk/29876/}

\bibitem[\protect\citeauthoryear{{von Braun-Bates}, {Winther}, {Alonso}  \&
  {Devriendt}}{{von Braun-Bates} et~al.}{2017}]{vonBraun-Bates2017}
{von Braun-Bates} F.,  {Winther} H.~A.,  {Alonso} D.,   {Devriendt} J.,  2017,
  \mn@doi [\jcap] {10.1088/1475-7516/2017/03/012}, \href
  {http://adsabs.harvard.edu/abs/2017JCAP...03..012V} {3, 012}

\makeatother
\end{thebibliography}
